# Thermodynamic characterization of the ($CO_2$ + $O_2$) binary system for the development of models for CCS processes: Accurate experimental (p, ρ, T) data and virial coefficients

Daniel Lozano-Martín[1], David Vega-Maza[1], M. Carmen Martín[1], Dirk Tuma[2], and César R. Chamorro[1].

[1] Grupo de Termodinámica y Calibración (TERMOCAL), Research Institute on Bioeconomy, Escuela de Ingenierías Industriales, Universidad de Valladolid, Paseo del Cauce, 59, E-47011 Valladolid, Spain.

[2] BAM Bundesanstalt für Materialforschung und -prüfung, D-12200 Berlin, Germany.

**Abstract**

Continuing our study on ($CO_2$ + $O_2$) mixtures, this work reports new experimental ($p$, $\rho$, $T$) data for two oxygen-rich mixtures with mole fractions $x(O_2)$ = (0.50 and 0.75) mol·mol$^{-1}$, in the temperature range $T$ = (250 to 375) K and pressure range $p$ = (0.5 to 20) MPa, using a single-sinker densimeter. Experimental density data were compared to two well established equation-of-state models: EOS-CG and GERG-2008. In the $p$, $T$-range investigated, the EOS-CG gave a better reproduction for the equimolar mixture ($x(O_2)$ = 0.5), whereas the GERG-2008 performed significantly better for the oxygen-rich mixture ($x(O_2)$ = 0.75). The EOS-CG generally overestimates the density, while the GERG-2008 underestimates it. This complete set of new experimental data, together with previous measurements, are used to calculate the virial coefficients $B(T, x)$ and $C(T, x)$, as well as the second interaction virial coefficient $B_{12}(T)$ for the ($CO_2$ + $O_2$) system.

* Corresponding author e-mail: cescha@eii.uva.es. Tel.: +34 983423756. Fax: +34 983423363

## 1. Introduction

High-accuracy density data are of great relevance for the development of reliable equations of state. In part one of this study [1], we reported accurate density measurements for three binary mixtures of carbon dioxide with oxygen (amount-of-substance fraction $x(O_2)$ = 0.05, 0.10, 0.20) in the temperature range $T$ = (250 to 375) K and maximum pressures up to $p$ = 13 MPa, together with the corresponding calculations using the two equation-of-state models GERG-2008 [2] and EOS-CG [3]. It could be observed that the GERG-2008 EoS fitted the experimental data within its claimed uncertainty in density (1 %) only for the mixture with the lowest oxygen content (amount-of-substance fraction $x(O_2)$ = 0.05). When the oxygen content increased ($x(O_2)$ = 0.10, 0.20), the deviations increased above the claimed uncertainty of the EoS and became more visible at lower temperatures and higher pressures. These deviations could be as high as 4.4 % for the mixture with $x(O_2)$ = 0.10, and 6.6 % for the mixture with $x(O_2)$ = 0.20. The deviations always had a positive value, i.e., the GERG-2008 underestimates the density of ($CO_2$ + $O_2$) mixtures, particularly for mixtures with a high oxygen content at high pressures and low temperatures. Another result of that study was that the EOS-CG performed much better in processing these density data. The relative deviations of the experimental density data from the EOS-CG remained within the claimed uncertainty of the equation of state (1 %) for all the 162 experimental points, with three exceptions only, namely at ($x(O_2)$ = 0.10, $T$ = 293.15 K, $p$ = 6.0 MPa), ($x(O_2)$ = 0.20, $T$ = 300 K, $p$ = 11.0 MPa), and ($x(O_2)$ = 0.20, $T$ = 300 K, $p$ = 12.2 MPa), where the relative deviation increased up to 1.2 %, −2.0 %, and −3.2 %, respectively.

In order to complete the characterization of the ($CO_2$ + $O_2$) binary mixture over the entire composition range, accurate density measurements for two new binary mixtures with higher oxygen content ($x(O_2)$ = 0.50, 0.75) are presented in this work. Measurements were performed at temperatures between (250 and 375) K and pressures up to 20 MPa using a single-sinker densimeter with magnetic suspension coupling, which is the same experimental technique used in the previous work. In order to achieve the highest accuracy in composition, the binary mixtures for this investigation were also prepared gravimetrically according to the ISO 6142-1 [4], a method that qualifies for the production of reference materials. The experimental results were compared with the GERG-2008 equation of state as well as with the more specific EOS-CG.

The complete set of density data for the binary system ($CO_2 + O_2$) presented in this work, and in the previous work [1], covers a wide range of temperature (from $T$ = 250 K to $T$ = 375 K), pressure (up to $p$ = 20 MPa), and composition (from $x(O_2)$ = 0.05 to $x(O_2)$ = 0.75). This complete set of new experimental data from both studies is used in this work to calculate the virial coefficients, $B(T, x)$ and $C(T, x)$, as well as the second interaction virial coefficient $B_{12}(T)$ for the ($CO_2 + O_2$) binary mixture.

The characterization of the binary system ($CO_2 + O_2$) is relevant not only for the development of accurate models for Carbon Capture and Storage (CCS) processes and for the modeling of combustion processes, but also for the improvement of the models used when dealing with natural gas and natural-gas-related mixtures. The deviations of the theoretical models from the actual values of the thermodynamic properties of the mixtures have relevant implications in the design and operation of processes, and in the transfer and pricing of products.

## 2. Experimental

### 2.1. Mixture preparation

Two ($CO_2 + O_2$) binary mixtures were prepared at the Federal Institute for Materials Research and Testing (Bundesanstalt für Materialforschung und -prüfung, BAM) in Berlin, Germany, according to the ISO 6142-1 [4].

Purity, supplier, molar mass, and critical parameters of the pure compounds (obtained from the reference equations of state for carbon dioxide [5] and oxygen [6]) are given in Table 1. The cylinder identifiers (BAM reference gas mixture G 033), the gravimetric composition, and the corresponding expanded uncertainty ($k$ = 2) of the mixtures are set out in Table 2. The prepared mixtures were supplied in aluminum cylinders with a volume of 10 $dm^3$. The entire mixture preparation procedure was executed in the same way as described in part one of this study [1]. Carbon dioxide and oxygen were used without further purification, but information on impurities from the specification by the supplier was considered in the mixture preparation.

Here, the following gas portions were determined that resulted in the final pressures:

Cylinder 1009-180717 ($x(O_2) = 0.50$): 1331.724 g $CO_2$ 968.171 g $O_2$ $p$ = 13.5 MPa

Cylinder 1099-180717 ($x(O_2) = 0.75$): 667.386 g $CO_2$ 1455.068 g $O_2$ $p$ = 13.6 MPa

The cylinders were also validated at BAM by gas chromatography using the same procedures as before [1]. Due to the composition of the mixtures, unlike part one, the analyzed compound was $CO_2$ in both cases. The results of the GC analysis and the composition of the mixtures used for validation are given in Table 3.

### 2.2. Equipment description

As in the previous work [1], the ($p$, $\rho$, $T$) data were measured using a single-sinker magnetic suspension densimeter (SSMSD) especially designed for density measurements of pure gases and gaseous mixtures. Details of the equipment and measurement procedure have previously been described by Chamorro et al. [7], Mondéjar et al. [8], and Lozano-Martín et al. [9]. This method, originally developed by Brachthäuser et al. [10] and improved by Klimeck et al. [11], operates on the Archimedes principle. A magnetic suspension coupling system allows the buoyancy force on a sinker immersed in the gas to be determined, so accurate density measurements of fluids over wide temperature and pressure ranges can be obtained. The setting of the equipment and the installed devices can be found in [12].

### 2.3. Density measurement procedure

The procedures for measuring densities are the same as in part one of this study [1], where the corresponding details can be found. Additional details of the measurement procedure in SSMSD are presented by Mondéjar et al. [8] and Lozano-Martín et al. [9] for our equipment and by McLinden [13] and Richter and Kleinrahm [14] on general aspects. In a simplified way, the density of the fluid can be calculated from Eq. (1):

$$\rho_{\mathrm{fluid}} = \frac{m_{\mathrm{s0}} - m_{\mathrm{sf}}}{V_{\mathrm{s}}(T,p)} \quad (1)$$

where the difference between the result of weighing the sinker in a vacuum $m_{s0}$, and in the pressurized fluid $m_{sf}$ is related to the buoyancy force exerted on the sinker. It is determined using a high-precision microbalance. $V_s$ ($T$, $p$) is the volume of the sinker immersed in the fluid, whose dependence on temperature and pressure is accurately known [8].

As explained in the previous work [1], the calibration of the balance and the correction due to the force transmission error (FTE) were considered [9][15], while the mass-based magnetic susceptibilities $\chi_s$ for the two ($CO_2$ + $O_2$) binary mixtures studied in this work were estimated using the additive law proposed by Bitter [16].

**2.4. Experimental procedure**

Experimental density data for the two ($CO_2$ + $O_2$) binary mixtures ($x(O_2)$ = 0.50 and 0.75) were obtained at temperatures of (250, 260, 275, 293.15, 300, 325, 350, and 375) K and pressures up to 20 MPa, always in the gas phase, whose limits are previously calculated with the EOS-CG [3] in order to stay always at pressures well below the saturation curve for each temperature. In the same way as applied in part one, the pressure was reduced in 1 MPa steps from the highest measured pressure to 1 MPa for each isotherm during a measurement campaign. The coordinates in Figure 1 show the recorded data in a $p$, $T$-diagram, together with the saturation curve for the mixture calculated with the EOS-CG [3]. Additionally, the $p$, $T$-range of applicability of the EOS-CG and the relevant area of interest for CCS applications are also indicated in the two plots of Figure 1.

Each individual coordinate was evaluated from thirty repeated measurements of each single ($p$, $\rho$, $T$) point and the last ten values are used to obtain the mean value. The balance calibration factor $\alpha$ is obtained right before and after every single point, and the apparatus-specific effect $\Phi_0$ is determined at the end of every single isotherm.

To minimize sorption effects inside the measuring cell, which may cause errors up to 0.1 % in density, the measuring cell was evacuated and flushed several times with fresh mixture before the isotherm is started, as recommended by Richter and Kleinrahm [14]. The residence time of the mixture in the cell never exceeded 40 hours. In this study, specific sorption tests for this particular mixture were performed in the same way as in previous works [12], [17]–[27]. Continuous density measurements on the same state point were recorded over 48 hours to detect any drifting. The results showed that the difference observed in the trend of the relative deviation in density between the measured and the calculated densities, using GERG-2008 EoS, between the first and last measurements of one campaign are one order of magnitude lower than the density uncertainty achievable with the equipment. A measurement with fresh mixture executed immediately afterwards, for the same temperature and pressure, reproduced the density value with a deviation of one order of magnitude lower than the density uncertainty of the equipment. Consequently, residual errors due to sorption effects are not appreciable with the experimental technique, and it should be considered that they are already included in the measurement uncertainty of the density and in the uncertainty in composition.

### 2.5. Uncertainty of the measurements

A detailed analysis of the uncertainties of the measurements involved in this experimental procedure was reported in previous works [8][9]. The quantities which contribute to the uncertainty of the measurements in this study are as follows:

The expanded uncertainty in temperature ($k$ = 2) is less than 4 mK. The pressure uncertainty depends on the range and is given by Eq. (2) and Eq. (3) for the (3 to 20) MPa and (0 to 3) MPa transducers, respectively. The expanded uncertainty ($k$ = 2) in pressure is, in both cases less than 0.005 MPa.

$$U(p)/\mathrm{MPa} = 75\cdot10^{-6}\cdot p/\mathrm{MPa} + 3.5\cdot10^{-3} \tag{2}$$

$$U(p)/\mathrm{MPa} = 60\cdot10^{-6}\cdot p/\mathrm{MPa} + 1.7\cdot10^{-3} \tag{3}$$

The uncertainty of density data for the two ($CO_2$ + $O_2$) binary mixtures investigated, corrected by both the apparatus-specific and the fluid-specific FTE effects, $U(\rho_{\mathrm{fluid}})$, is evaluated according to the methods proposed in the Guide to the Expression of Uncertainty in Measurement (GUM) [28]. Eq. (4) is the working equation used in this study, where $\chi_s$ stands for the mass-based magnetic susceptibility.

$U(\rho)$/kg·m$^{-3}$= 2.5·10$^{4}$·$\chi_s$/m$^3$·kg$^{-1}$ + 1.1·10$^{-4}$·$\rho$/kg·m$^{-3}$ + 2.3·10$^{-2}$ (4)

The resulting working equation (Eq. (5)) to calculate the overall expanded uncertainty in density $U_T(\rho)$ ($k$ = 2) includes uncertainties of density, temperature, pressure, and composition of the mixture.

$$U_T(\rho) = 2 \cdot \left[ u(\rho)^2 + \left( \left( \frac{\partial \rho}{\partial p} \right)_{T,x} \cdot u(p) \right)^2 + \left( \left( \frac{\partial \rho}{\partial T} \right)_{p,x} \cdot u(T) \right)^2 + \sum_i \left( \left( \frac{\partial \rho}{\partial x_i} \right)_{T,p,x_j \neq x_i} \cdot u(x_i) \right)^2 \right]^{0.5} \quad (5)$$

In Eq. (5), $p$ is the pressure, $T$ is the temperature, and $x_i$ is the amount-of-substance (mole) fraction of each mixture component. Partial derivatives were calculated from the GERG-2008 EoS using the REFPROP software [29].

The individual contributions of density, temperature, pressure, and composition to the overall uncertainty in density for the three studied ($CO_2$ + $O_2$) binary mixtures are given in Table 4.

## 3. Experimental results

Tables 5 and 6 show the 274 experimental ($p$, $\rho$, $T$) data measured for the two ($CO_2$ + $O_2$) binary mixtures. The temperature, pressure, and density of each measured point were calculated as the arithmetic mean of the last ten consecutive measurements of a series of thirty. Tables 5 and 6 also show the expanded uncertainty in density $U(\rho_{exp})$ ($k$ = 2), calculated by Eq. (4) and expressed in absolute density units and as a percentage of the measured density.

The experimental data were compared to the corresponding densities calculated from the two equations of state GERG-2008 and EOS-CG, using the REFPROP [29] and TREND 4.0 [30] software. Relative deviations of the experimental densities from the corresponding EoS-values are included in Tables 5 and 6 and are shown in Figures 2 and 3.

The densities of the experimental points recorded in this work range from $\rho$ = 9.294 kg·m$^{-3}$ ($T$ = 250 K, $p$ = 0.5 MPa, $x(O_2)$ = 0.50) to $\rho$ = 536.87 kg·m$^{-3}$ ($T$ = 275 K, $p$ = 18.8 MPa, $x(O_2)$ = 0.50).

Note that analogously to part one of this study, there is a correction applied due to the fluid-specific effect originating from the content of a paramagnetic fluid, namely oxygen. This correction can be applied thanks to the estimation of the apparatus-specific constant $\varepsilon_\rho$ of the fluid-specific effect in a previous work [9]. The contribution of this correction is much higher for the mixtures measured in this work, as the oxygen content is much higher. In fact, the correction due to the fluid-specific effect can be as high as 6.701 kg·m$^{-3}$ in absolute value (1.49 % relative value), at the highest density of $\rho$ = 448.577 kg·m$^{-3}$; or as high as 2.13 % in relative value (0.368 kg·m$^{-3}$ absolute value), at the lowest density of $\rho$ = 17.245 kg·m$^{-3}$, for the mixture with the higher oxygen content (0.25 $CO_2$ + 0.75 $O_2$) at $T$ = 250 K.

## 4. Discussion of the results

### 4.1. Relative deviation of the experimental data from the reference equations of state

The plot in Figure 2 shows the relative deviations of the experimentally determined density data of the (0.50 $CO_2$ + 0.50 $O_2$) mixture from the corresponding density data calculated by the GERG-2008 (a) and the EOS-CG (b) models, respectively. In the same way, Figure 3 shows the deviations for the (0.25 $CO_2$ + 0.75 $O_2$) mixture.

Both equations of state claim an uncertainty in density of 1.0 % for mixtures of $CO_2$ and $O_2$ over the temperature range from (250 to 450) K and at pressures up to 35 MPa. The estimated uncertainty of experimental density data ranges from 0.019 % for $T$ = 275 K, $p$ = 18.8 MPa ($\rho$ = 536.87 kg·m$^{-3}$) for the (0.50 $CO_2$ + 0.50 $O_2$) mixture to 0.377 % for $T$ = 375 K, $p$ = 1.0 MPa ($\rho$ = 11.262 kg·m$^{-3}$) for the (0.25 $CO_2$ + 0.75 $O_2$) mixture. A slightly bigger relative uncertainty of 0.438 % can be found in a single point for the (0.50 $CO_2$ + 0.50 $O_2$) mixture at $T$ = 250 K, but this value is due to the fact that this point is the only one that has been measured at the pressure of $p$ = 0.5 MPa, with a density as low as 9.294 kg·m$^{-3}$).

The relative deviations of the experimental density data from the corresponding data of GERG-2008 (Figures 2 (a) and 3 (a)) are larger for the mixture with lower oxygen content, i.e., the (0.50 $CO_2$ + 0.50 $O_2$) mixture. Here, 55 of the 123 experimental points deviate more than the claimed uncertainty of the equation of state. This behavior emerges for all the measured temperatures except the two highest at $T$ = 350 K and $T$ = 375 K, respectively. The relative deviations can be as large as 3.90 %. For the mixture with the higher oxygen content, i.e., the (0.25 $CO_2$ + 0.75 $O_2$) mixture, only 2 of the 151 measured points (at $T$ = 250 K, and $p$ = 10.0 MPa and $p$ = 11.0 MPa) deviate slightly more (1.05 %) than the claimed uncertainty of the EoS. These two points are close to the saturation curve, as can be seen in Figure 1(b). Almost all the deviations have a positive value which means that the GERG-2008 EoS underestimates the density of ($CO_2$ + $O_2$) mixtures. Further, the course of the deviation is not monotonous. With increasing pressures, the curves pass a maximum and the deviation diminishes towards the maximum pressure of 20 MPa. This maximum is located at approximately 10 MPa for the (0.25 $CO_2$ + 0.75 $O_2$) mixture; whereas, for the (0.50 $CO_2$ + 0.50 $O_2$) mixture, higher temperatures shift this maximum deviation towards higher pressures.

In contrast to the GERG-2008, the relative deviations of the experimental density data from the corresponding data of the EOS-CG increase as the oxygen content in the mixture increases. For the (0.50 $CO_2$ + 0.50 $O_2$) mixture, 23 of the 123 experimental points deviate more than the claimed uncertainty of the EOS-CG (Figure 3 (b)). This occurs for temperatures of 275 K, 293.15 K, and 300 K and pressures higher than 10 MPa. The relative deviations reach a maximum of –2.95 %. For the mixture with the higher oxygen content, i.e., the (0.25 $CO_2$ + 0.75 $O_2$) mixture, 33 of the 151 measured points deviate more than the claimed uncertainty of the EOS-CG (Figure 4 (b)). This region is entered for pressures over 9 MPa at $T$ = 250 K, over 10 MPa at $T$ = 260 K, over 12 MPa at $T$ = 275 K, over 15 MPa at $T$ = 293.15 K, and only over 18 MPa at $T$ = 300 K. Here, the relative deviations can be as large as –2.64 %. The larger deviations always have a negative value, i.e., the EOS-CG overestimates the density of ($CO_2$ + $O_2$), being more pronounced for mixtures with high oxygen content, at high pressures and low temperatures, and the slope of the curve displays a minimum. In reverse analogy to the GERG-2008, higher temperatures move this minimum towards higher pressures.

### 4.2 Virial coefficients

The virial coefficients for the five ($CO_2$ + $O_2$) mixtures, the three measured in the previous work ($x(O_2)$ = 0.05, 0.10, and 0.20) [1] and the two measured in this work ($x(O_2)$ = 0.50 and 0.75), were calculated by fitting the experimental density data to the virial EoS:

$$\frac{p}{RT} = \sum_{k=1}^{N} \frac{B_k}{M^k} \rho_{\mathrm{exp}}^k \qquad (6)$$

where $p$ is the pressure, $T$ is the temperature, $R$ is the molar gas constant, $\rho_{\mathrm{exp}}$ is the experimental density, $M$ is the molar mass, and $B_k$ with $k$ = 1, 2, … ($B_1$ = 1) are the second, third,… virial coefficients, respectively. The method proposed by Cristancho *et al.* [31] was used to determine the number of terms at which the virial EoS must be truncated and the maximum density $\rho_{\mathrm{max}}$ of the experimental points used to fit this equation. The procedure, described in detail in a previous work [23], consists of two consecutive fits. Both fits were executed using a least-squares fitting method implemented in MATLAB software [32].

The first fit determines the number of virial coefficients $N$ needed for the best representation of the experimental data, and the maximum density $\rho_{\mathrm{max}}$ for which the fit gives a satisfying result. This is done through the determination of the apparent molar mass $M$ of the mixture as a parameter of the virial EoS, while varying the number of coefficients of the virial-EoS $N$ and the range of the experimental data sets until the obtained value of $M$ is as close as possible to the accepted reference value of $M$ for each mixture. The first fit is performed under the following conditions: that the number of experimental points to be fitted is higher than $2 \cdot N$, the standard uncertainty of all the virial coefficients is lower than their own value (then, the parameters are significant), and the root mean square of the residuals is within the expanded ($k$ = 2) experimental uncertainty in density.

The first fit yielded that the closest values of $M$ are obtained with a third order ($N$ = 3) of the virial EoS for all the compositions and temperatures. The value of $\rho_{\mathrm{max}}$ is between (117.407 and 153.813) $kg \cdot m^{-3}$ ($p \approx 7$ to 8 MPa) for the (0.95 $CO_2$ + 0.05 $O_2$) mixture, between (123.409 and 231.402) $kg \cdot m^{-3}$ ($p \approx 5$ to 7 MPa) for the (0.90 $CO_2$ + 0.10 $O_2$) mixture, between (118.408 and 264.364) $kg \cdot m^{-3}$ ($p \approx 7$ to 9 MPa) for the (0.80 $CO_2$ + 0.20 $O_2$) mixture, between (146.474 and 426.654) $kg \cdot m^{-3}$ ($p \approx 11$ to 19 MPa) for the (0.50 $CO_2$ + 0.50 $O_2$) mixture, and lastly between (127.886 and 332.575) $kg \cdot m^{-3}$ ($p \approx 11$ to 15 MPa) for the (0.25 $CO_2$ + 0.75 $O_2$) mixture. For some isotherms, it was not possible to obtain a regression complying with the three conditions indicated above for all studied compositions. The virial coefficients at the lowest temperatures, $T$ = (250 and 260) K, were only obtained for the mixture with the higher oxygen content (0.25 $CO_2$ + 0.75 $O_2$), and the virial coefficients at $T$ = 275 K were only obtained for the two mixtures with higher oxygen content, ($0.50\ CO_2 + 0.50\ O_2$) and ($0.25\ CO_2 + 0.75\ O_2$). This is because these relatively low-temperature isotherms fall just below the phase envelopes for the mixtures with lower oxygen content (and thus, high carbon dioxide content) and the explored experimental range includes no more than 5 points to a rather low $\rho_{max}$ of 109.732 kg·m$^{-3}$ (at $T$ = 275 K, $p$ = 3.94 MPa for the ($0.95\ CO_2 + 0.05\ O_2$) mixture)

The second fit calculates the values of the corresponding virial coefficients, using the values of $N$ and $\rho_{max}$ obtained in the first fit. The final calculus is performed with the value of $M$ fixed to the accepted reference value for each mixture. The results for the second $B(T, x)$ and third $C(T, x)$ virial coefficients are reported in Table 7 for all the five binary ($CO_2 + O_2$) mixtures of this work, together with their uncertainty determined by the Monte Carlo method [33]. Three points were treated as outliers and they are reported neither in Table 7 nor in Figure 4, namely those at $T$ = 375 K for the ($0.95\ CO_2 + 0.05\ O_2$) mixture and at $T$ = 350 K for the ($0.95\ CO_2 + 0.05\ O_2$) and ($0.90\ CO_2 + 0.10\ O_2$) mixtures.

Table 7 and Figures 4 and 5 also report the second interaction virial coefficients $B_{12}$ obtained from the second virial coefficients using the reference EoS of pure carbon dioxide $B_{11}$ [5], pure oxygen $B_{22}$ [6], and the expression:

$$B(T,x) = x_1^2 B_{11}(T) + 2x_1 x_2 B_{12}(T) + x_2^2 B_{22}(T) \quad (7)$$

where $x_1$ and $x_2$ are the mole fraction of carbon dioxide and oxygen, respectively. The expanded ($k$ = 2) uncertainty of $B_{12}$ has been determined applying the law of uncertainty propagation [28] to the uncertainties of the mixture's second virial coefficient $B$ from the Monte Carlo method, as described above, and the uncertainties of $B_{11}$ (0.5 cm$^3$·mol$^{-1}$) and $B_{22}$ (0.3 cm$^3$·mol$^{-1}$) from [34].

The second interaction virial coefficients $B_{12}$ for the binary ($CO_2 + O_2$) system investigated in our studies were fitted to:

$$B_{12} = N_0 + \frac{N_1}{T} \quad (8)$$

with the corresponding parameters reported in Table 8. The residuals of the fit to Eq. (8) are plotted in Figure 5b, with a root mean square of 2.3 %, within the 4.3 % average expanded ($k$ = 2) uncertainty of $B_{12}$.

Figure 4 and Table 7 depict $B_{12}$ as a function of the mole fraction of oxygen for each isotherm. As can be seen, the values computed from the EOS-CG show a dependence with the mixture composition, especially for the isotherms at the lowest temperatures; while the experimentally estimated values of $B_{12}$ have a flatter trend with the composition, as stated by the theory. Thus, at a higher content of oxygen, the experimental results of $B_{12}$ are less negative than the corresponding results evaluated from the EOS-CG. However, at the lowest mole fraction of oxygen, the behaviour is the opposite.

Figure 5a and Table 7 display the average value of $B_{12}$ from all the compositions as a function of the temperature and a comparison with the values given by the GERG-2008 model, the EOS-CG model, and available data in the literature. There is good agreement between the EOS-CG and the experimental $B_{12}$, with deviations within the $U(B_{12,\text{ exp}})$ = 4.3 %. Moreover, the experimental values of $B_{12}$ are consistent, considering their respective uncertainties, with the literature data of Martin et al. [35] ($U(B_{12,\text{Martin et al.}})$ = 3.3 %) and Gorski and Miller [36] ($U(B_{12,\text{Gorski and Miller}})$ = 0.7 %). Nevertheless, there are larger discrepancies with the data of Edwards and Roseveare [37] ($U(B_{12,\text{Edwards and Roseveare}})$ = 5.3 %) and a different trend with temperature is found concerning the data of Cottrell et al. [38] ($U(B_{12,\text{Cottrell et al.}})$ = 17.0 %). Notably, the deviations from applying the GERG-2008 EoS are significant. The values of $B_{12}$ obtained from the GERG-2008 EoS are significantly higher than those from the experiment. The deviations are one order of magnitude higher compared to those originating from the EOS-CG, ranging from 9 % up to 68 %, far beyond the $U(B_{12,\text{ exp}})$. This may be due to the fact that the binary system ($CO_2$ + $O_2$) in the GERG-2008 EoS is correlated only to binary vapour-liquid equilibrium data [39], while the EOS-CG also considers density [36][40][41], speed of sound [42], and interaction second virial coefficient data sets [35, 38] to regress the model. Moreover, the EOS-CG uses more accurate EoS for pure carbon dioxide [5] and oxygen [6], instead of the expressions used in the GERG-2008 EoS for the same substances [43][44].

### 4.3. General analysis of the joint data for the five different compositions

Considering the results presented in this work together with the results presented in the previous work [1] as a whole, we can say that, in general, the EOS-CG can reproduce the experimental density data better than the GERG-2008 EoS. This can be seen clearly in Figure 6 (a) and (b). This is not surprising, as the EOS-CG is specifically designed for this kind of mixtures, whereas the GERG-2008 is a more general approach. However, strictly speaking, the EOS-CG fits the experimental data within its claimed uncertainty only for the mixtures with the lower oxygen content ($x(O_2)$ = 0.05, 0.10 and 0.20). For the mixtures with higher oxygen content ($x(O_2)$ = 0.50 and 0.75), mainly at high pressures ($p$ > 10 MPa) and low temperatures ($T$ < 300 K), the EOS-CG cannot reliably reproduce the experimental data within its claimed uncertainty. Unexpectedly, the GERG-2008 can fit the experimental data within the uncertainty borders for the mixture with the highest oxygen content ($x(O_2)$ = 0.75) better than the EOS-CG, which presents deviations as high as –2.64 %. The virial equation of state, with the coefficients obtained in this work, can reproduce the experimental data with greater precision, as can be seen in Figure 6 (c), where the relative deviations of the experimental pressure from the pressure given by the virial equation of state are plotted as a function of density. The deviations are very small and only for the higher values of density, i.e., above 400 $kg \cdot m^{-3}$, are the deviations bigger than 0.4 %, with a maximum deviation of 1.3 % for the (0.50 $CO_2$ + 0.50 $O_2$) mixture and 3.4 % for the (0.80 $CO_2$ + 0.20 $O_2$) mixture.

Table 9 presents the statistical parameters of the relative deviation of the experimental data from the densities given by the EOS-CG and the GERG-2008, and from pressures given by the virial equation of state. The AAD of experimental data from the densities calculated by the EOS-CG amounts to 0.077 % for the (0.95 $CO_2$ + 0.05 $O_2$) mixture, 0.15 % for the (0.90 $CO_2$ + 0.10 $O_2$) mixture, 0.22 % for the (0.80 $CO_2$ + 0.20 $O_2$) mixture, 0.59 % for the (0.50 $CO_2$ + 0.50 $O_2$) mixture, and 0.66 % for the (0.25 $CO_2$ + 0.75 $O_2$) mixture. The corresponding AAD of experimental data from the densities calculated by the GERG-2008 are 0.29 %, 0.69 %, 1.3 %, 1.2 %, and 0.32 %. Only for the most oxygen-rich (0.25 $CO_2$ + 0.75 $O_2$) mixture is the AAD of the GERG-2008 smaller than that of the EOS-CG. The RMS of experimental data from the pressures given by the virial equation of state ranges between 0.035 % for the (0.95 $CO_2$ + 0.05 $O_2$) mixture and 0.47 % for the (0.80 $CO_2$ + 0.20 $O_2$) mixture.

The availability of data for ($CO_2$ + $O_2$) mixtures in the literature is limited to oxygen contents below $x(O_2)$ = 0.10 [40], [45][46][42]. For this reason, a direct comparison with these experimental data is difficult and will remain rather speculative. The data from these references were also processed to obtain the statistical parameters of the relative deviation of these experimental data from the densities calculated by the EOS-CG and the GERG-2008 and are given in Table 9.

Regarding the vapor-liquid equilibrium of the mixture, even when it is not the object of this work, it can be said, based on the analysis of available experimental data [1-6], that the ($CO_2$ + $O_2$) mixture is a Type I binary system according to the van Konynenburg and Scott classification [7]. This is what to expect from a binary system of molecules of similar size, non-dipolar moment and an average energy of the $CO_2$ - $O_2$ interaction of the same order of magnitude to the average energy of $CO_2$ - $CO_2$ and $O_2$ - $O_2$ interactions in the mixture [8].

## 5. Conclusions

New ($p$, $\rho$, $T$) high-precision experimental data for two binary mixtures of carbon dioxide and oxygen, with nominal compositions of (0.50 $CO_2$ + 0.50 $O_2$) and (0.25 $CO_2$ + 0.75 $O_2$), at temperatures between (250 and 375) K and pressures up to 20 MPa, are reported. The gravimetrically prepared mixtures were of reference quality and the experimental device used was a single-sinker densimeter with magnetic suspension coupling.

The new experimental data were compared to the corresponding densities calculated from the EOS-CG and the GERG-2008 equation-of-state models and rated by the uncertainty threshold of 1 %, which is applicable for both models. The results of all five mixtures ($x(O_2)$ = 0.05, 0.10, 0.20, 0.50, and 0.75) investigated were employed to determine virial coefficients.

For the two mixtures under study here, in the $p$, $T$-range investigated, the equimolar mixture ($x(O_2)$ = 0.5) was better reproduced by the EOS-CG but the GERG-2008 performed better on the mixture with the highest oxygen content ($x(O_2)$ = 0.75). The EOS-CG also gave better results for the three carbon dioxide-rich mixtures investigated in part one of this study ($x(O_2)$ = 0.05, 0.10, and 0.20) [1]. Generally, the GERG-2008 underestimated the density of all mixture compositions, while the deviations became larger towards lower temperatures and frequently surpassed the uncertainty threshold. In contrast to the GERG-2008, the relative deviations in density calculated by the EOS-CG did not show a distinct trend. There was no clear temperature dependence observed for the three carbon dioxide-rich mixtures and the values remained mostly within the uncertainty threshold over the entire *p*, *T*- range. For the other two mixtures, however, a tendency to underestimate the density was found, which became larger at lower temperatures and pressures > 10 MPa. Notably, the plot deviation versus pressure of both the GERG-2008 and the EOS-CG passed an extremum for several mixtures, namely the GERG-2008 on $x(O_2) = 0.20$ and both models on $x(O_2) = 0.50$ and 0.75.

The complete set of experimental data, which covers the entire composition range of the binary ($CO_2 + O_2$) mixture for the first time, together with the obtained virial coefficients, is an important tool to develop and improve the models and equations of state needed to design and operate processes with carbon dioxide and oxygen, such as CCS.

## Acknowledgments

The authors wish to thank for their support the Ministerio de Economía, Industria y Competitividad project ENE2017-88474-R and the Junta de Castilla y León project VA280P18.

## Figures

**Figure 1.** *p*, *T*-phase diagram showing the experimental points measured (•) and the calculated phase envelope (solid line) using the EOS-CG [3] for: (a) (0.50 $CO_2$ + 0.50 $O_2$) and (b) (0.25 $CO_2$ + 0.75 $O_2$) binary mixtures, respectively. The marked temperature and pressure ranges represent the range of the binary experimental data used for the development of the EOS-CG (red dashed line), the GERG-2008 EoS (blue dashed line), and the area of interest for the gas industry (black thin dashed line).

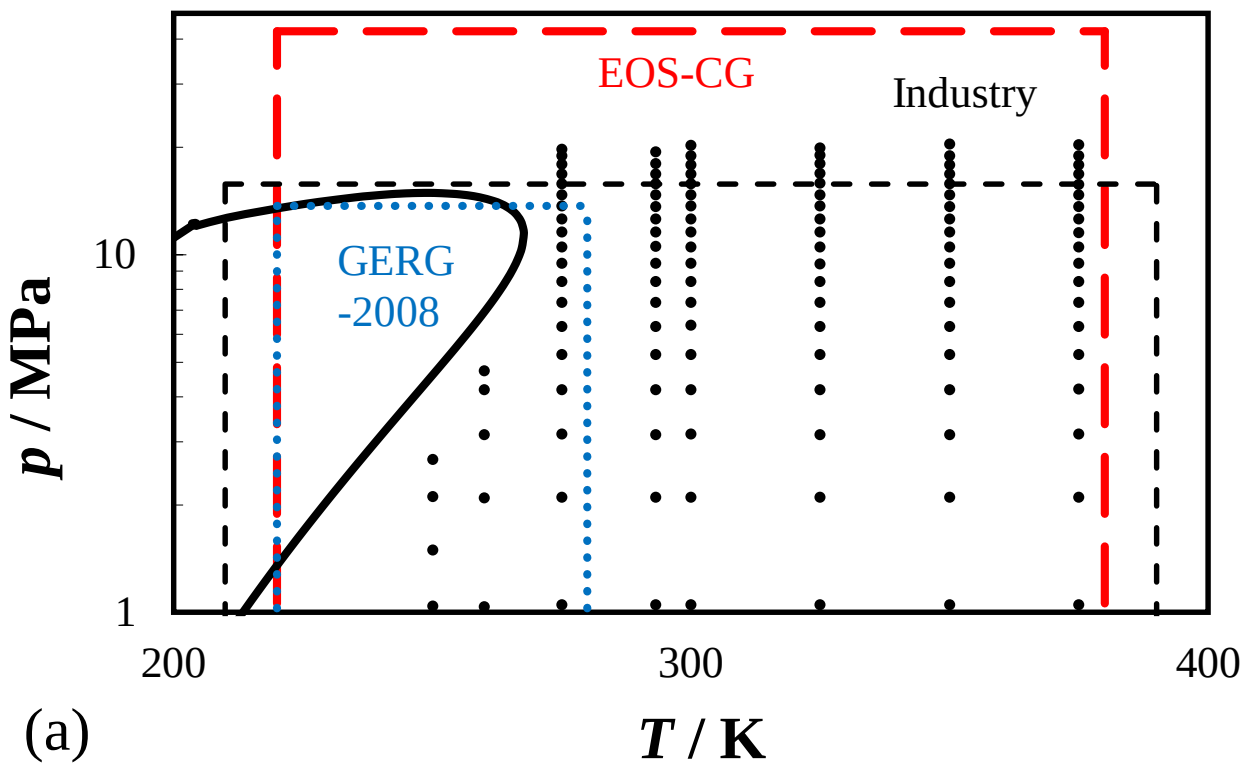

(a)

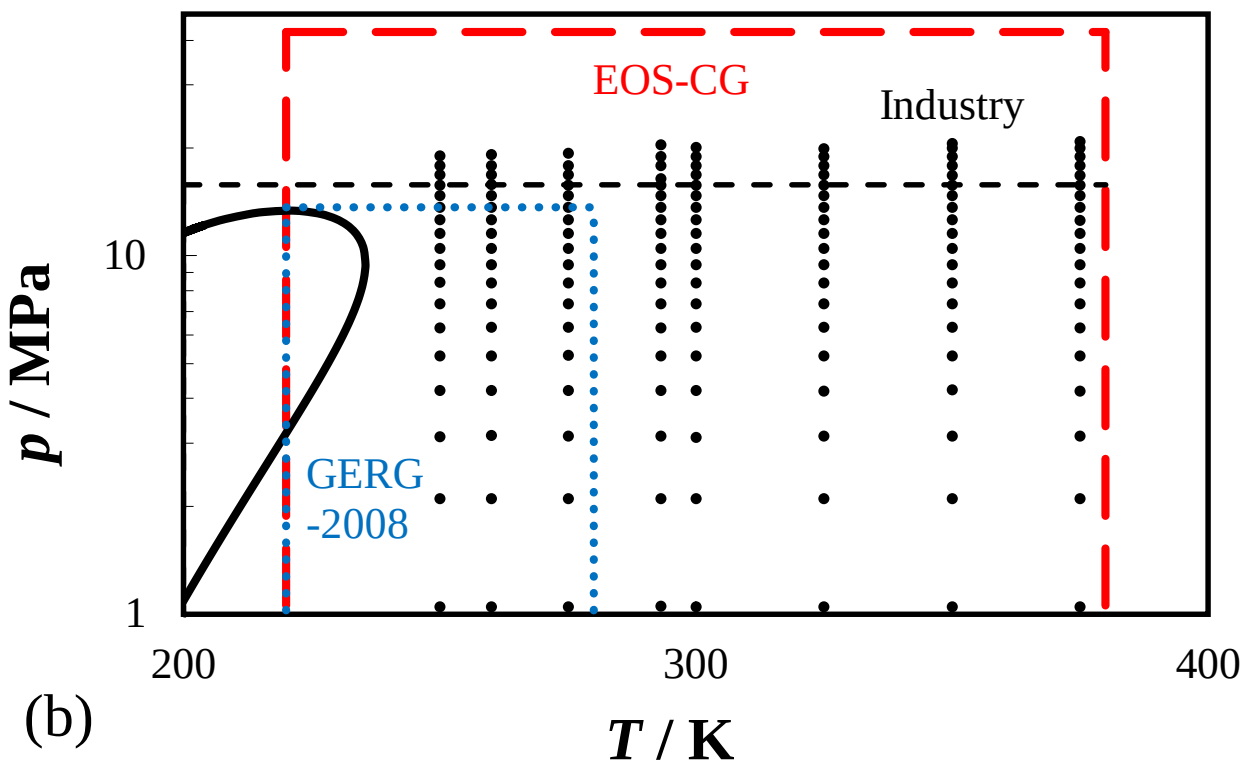

(b)

**Figure 2.** Relative deviations in density of the experimental ($p,\rho_{exp},T$) data of the binary (0.50 $CO_2$ + 0.50 $O_2$) mixture from density values calculated by the: (a) GERG-2008 [2], $\rho_{GERG}$, and (b) EOS-CG [3], $\rho_{CG}$, equations of state as a function of pressure for different temperatures: ◇ 250 K, △ 260 K, × 275 K, □ 293.15 K, ○ 300 K, + 325 K, ✳ 350 K, — 375 K. Dashed lines indicate the expanded ($k$ = 2) uncertainty of the corresponding EoS. Error bars on the 293.15 K data set indicate the expanded ($k$ = 2) uncertainty of the experimental density.

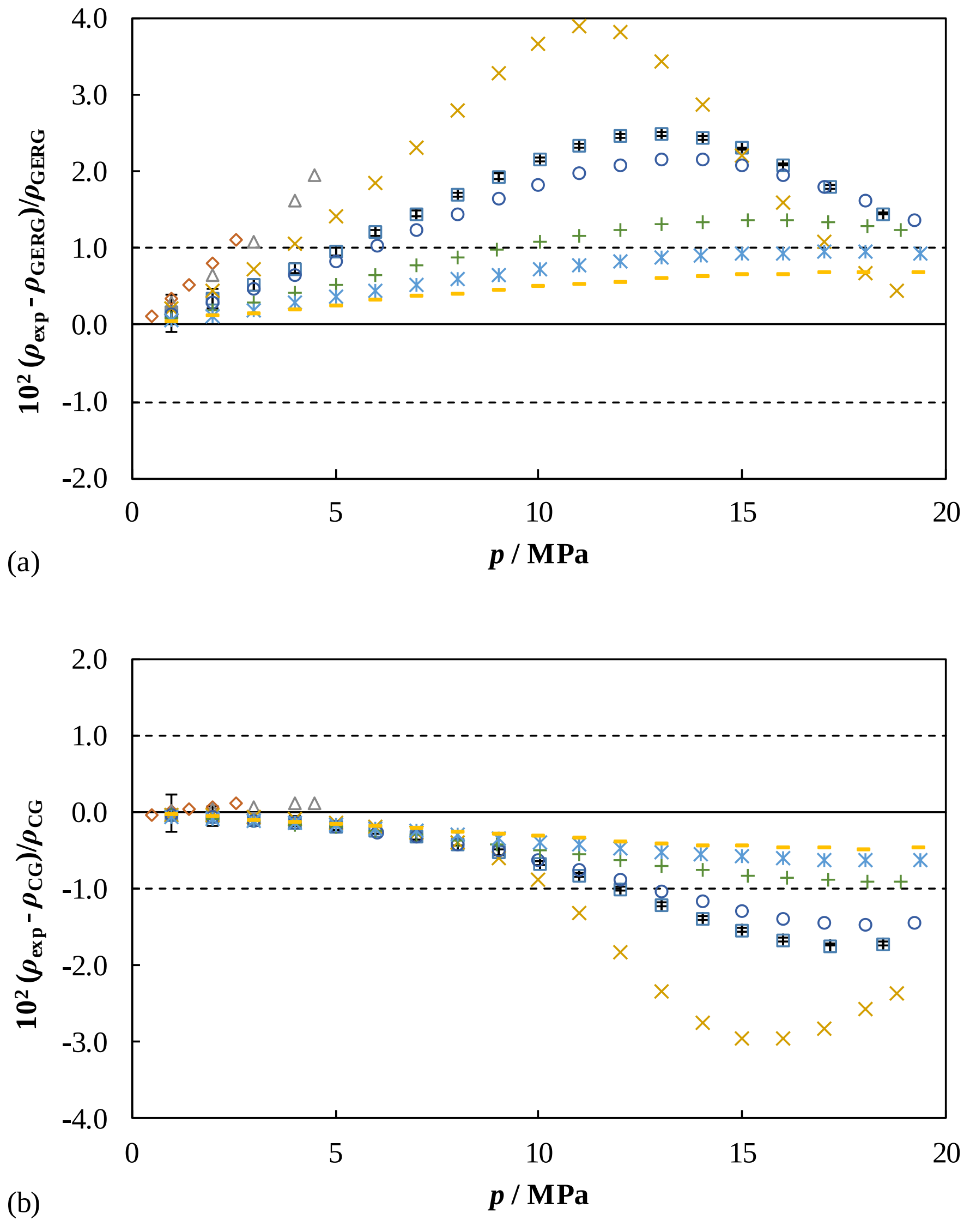

**Figure 3.** Relative deviations in density of the experimental ($p$,$\rho_{exp}$,$T$) data of the binary (0.25 $CO_2$ + 0.75 $O_2$) mixture from density values calculated by the: (a) GERG-2008 [2], $\rho_{GERG}$, and (b) EOS-CG [3], $\rho_{CG}$, equations of state as a function of pressure for different temperatures: ◇ 250 K, △ 260 K, × 275 K, □ 293.15 K, ○ 300 K, + 325 K, ✳ 350 K, — 375 K. Dashed lines indicate the expanded ($k$ = 2) uncertainty of the corresponding EoS. Error bars on the 293.15 K data set indicate the expanded ($k$ = 2) uncertainty of the experimental density.

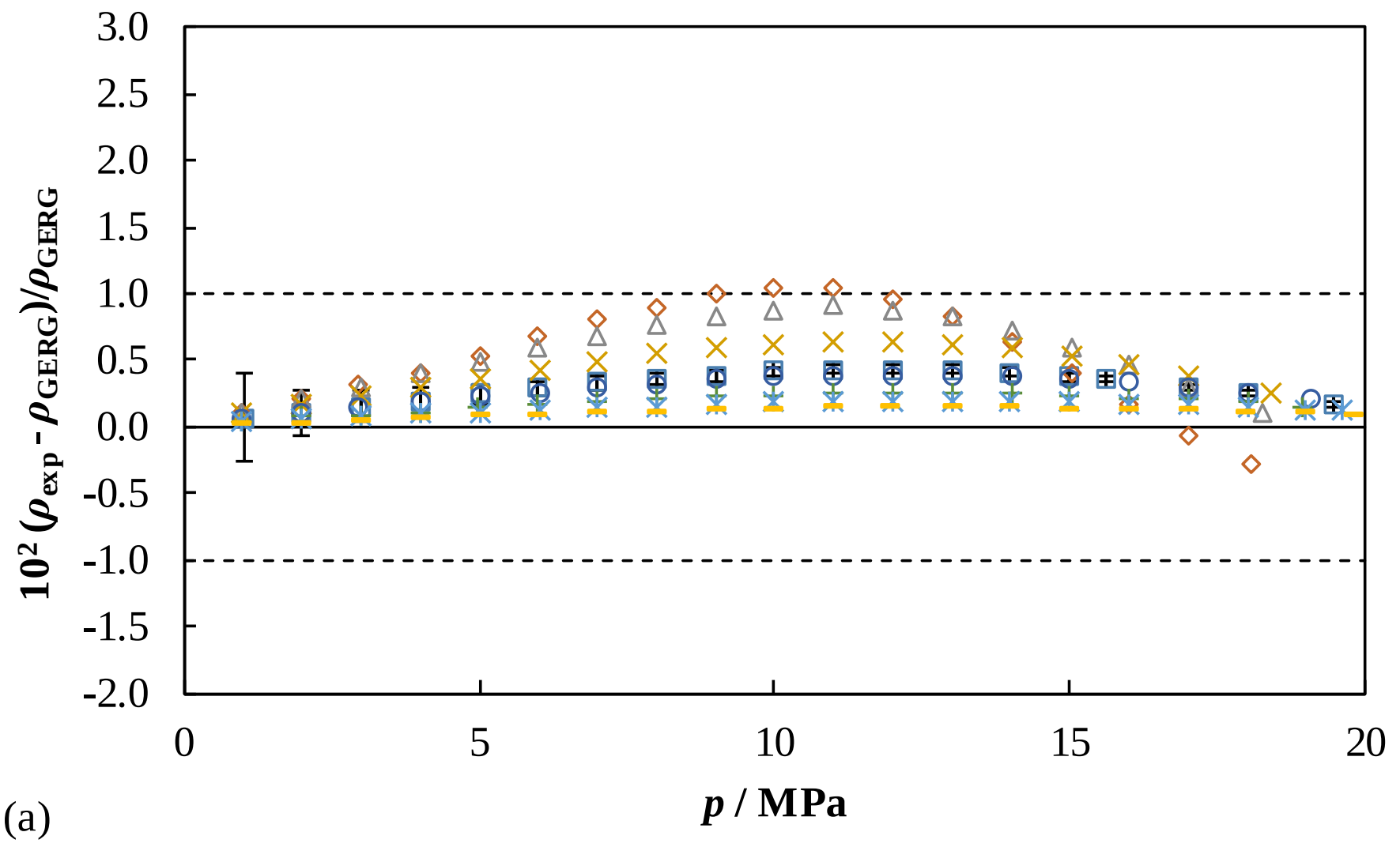

(a)

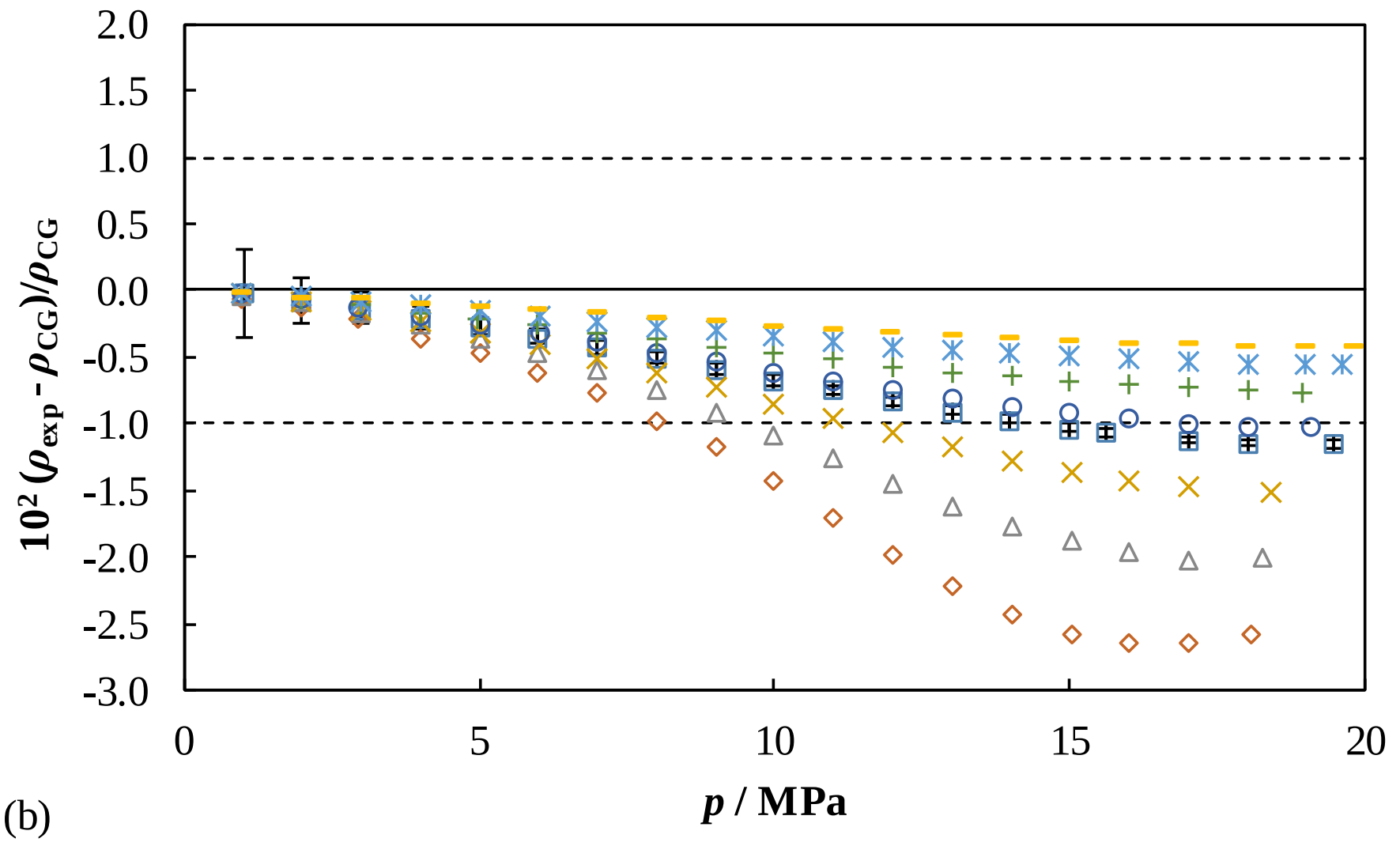

(b)

**Figure 4.** Second interaction virial coefficient $B_{12}(T)$ for the binary ($CO_2$ + $O_2$) system estimated from the experimental data as a function of the $O_2$ mole fraction, $x(O_2)$, at different temperatures: ◇ 250 K, △ 260 K, × 275 K, ○ 300 K, + 325 K, ✳ 350 K, − 375 K. The dashed lines represent the $B_{12}(T)$ values computed from the EOS-CG at the corresponding temperatures.

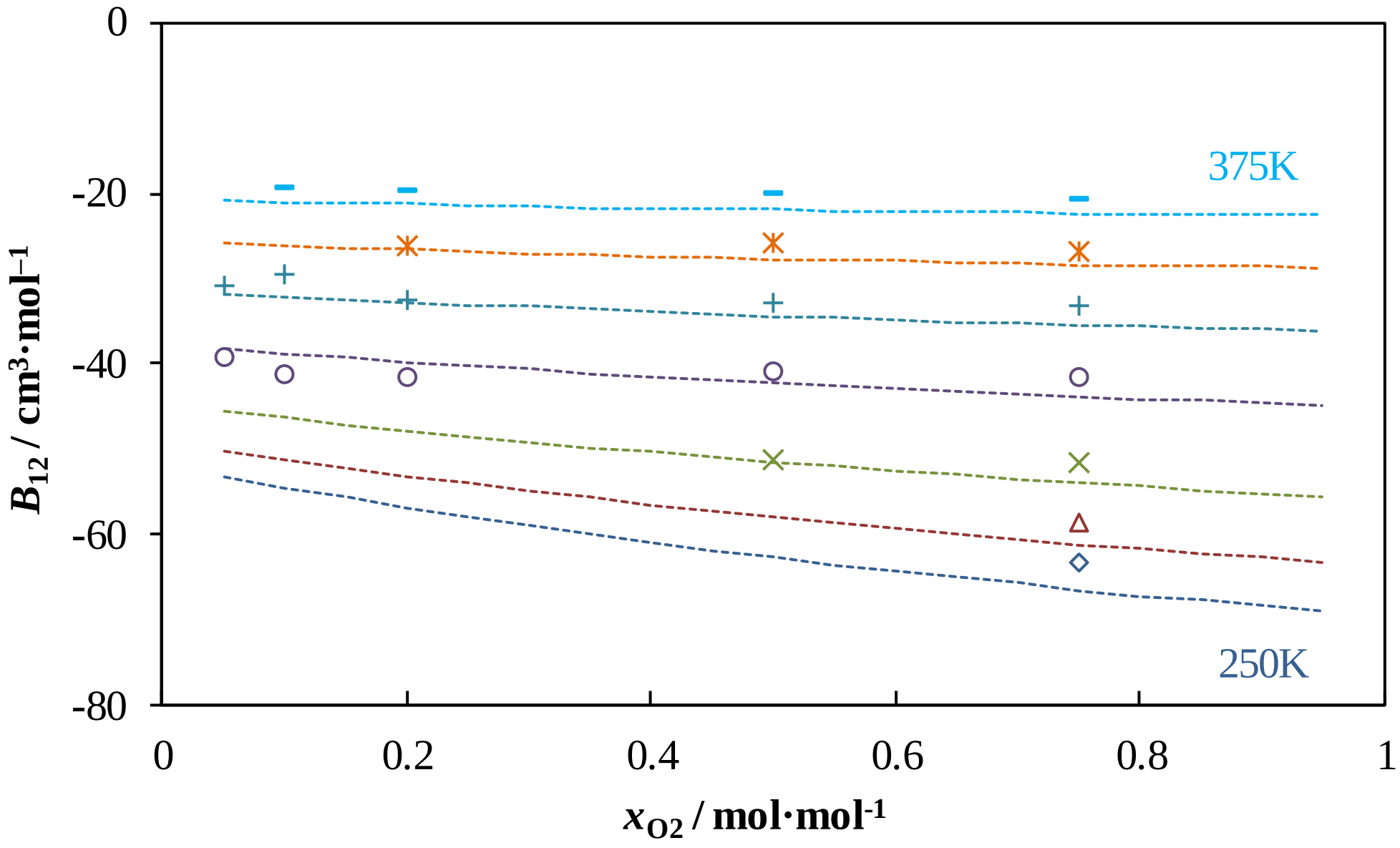

**Figure 5.** (a) Mean value of the second interaction virial coefficient $B_{12}(T)$ for the binary ($CO_2$ + $O_2$) system as a function of temperature from: × this work, △ Martin et al. [35], ◇ Edwards and Roseveare [37], □ Gorski and Miller [36], ○ Cottrell et al. [38], - - GERG-2008 EoS [2], ⋯ EOS-CG [3]. Error bars indicate the expanded ($k$ = 2) uncertainty of the estimated $B_{12}(T)$ values. The solid line represents the fit to equation (8) of the experimental data of this work. (b) Residuals of the fit to equation (8).

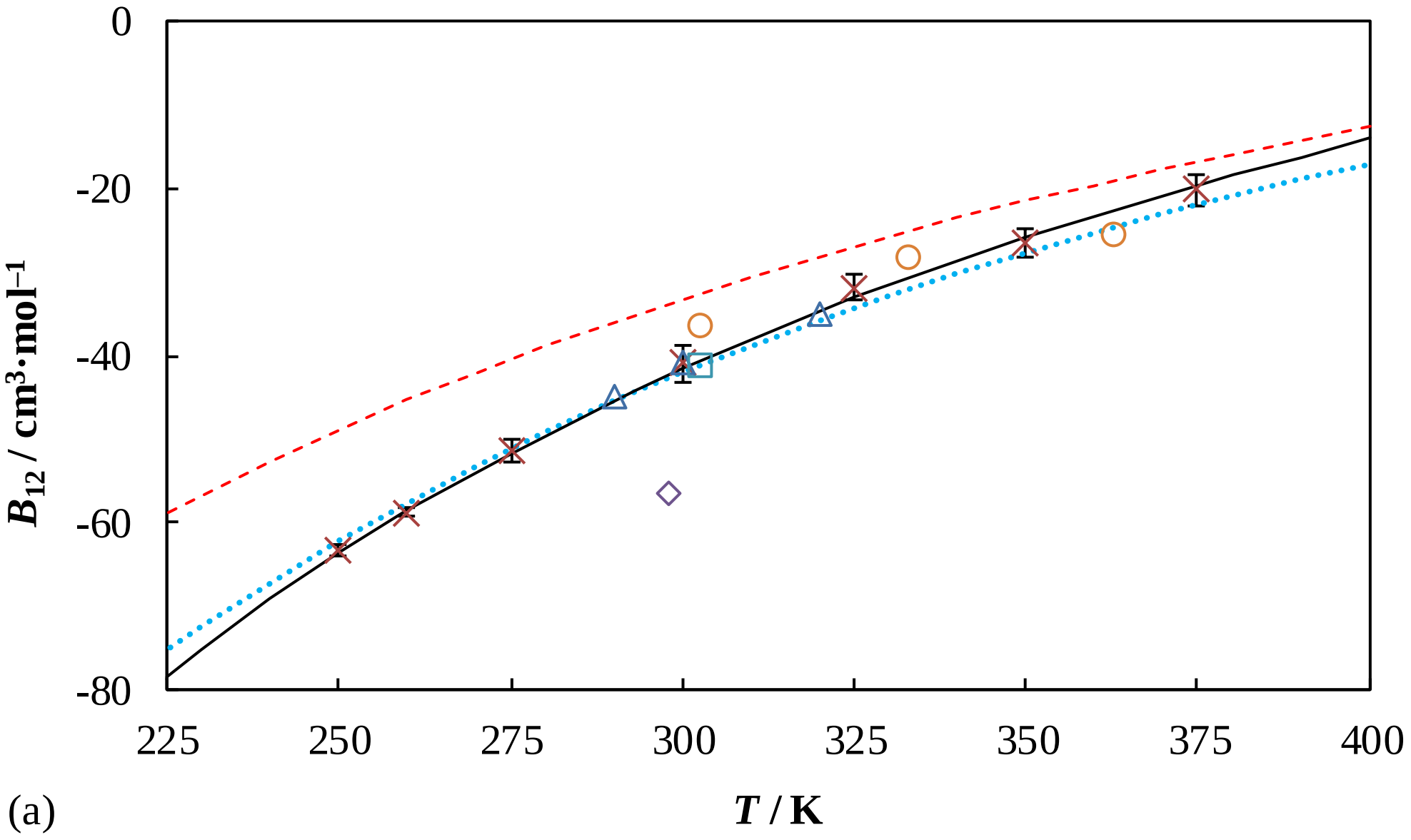

(a)

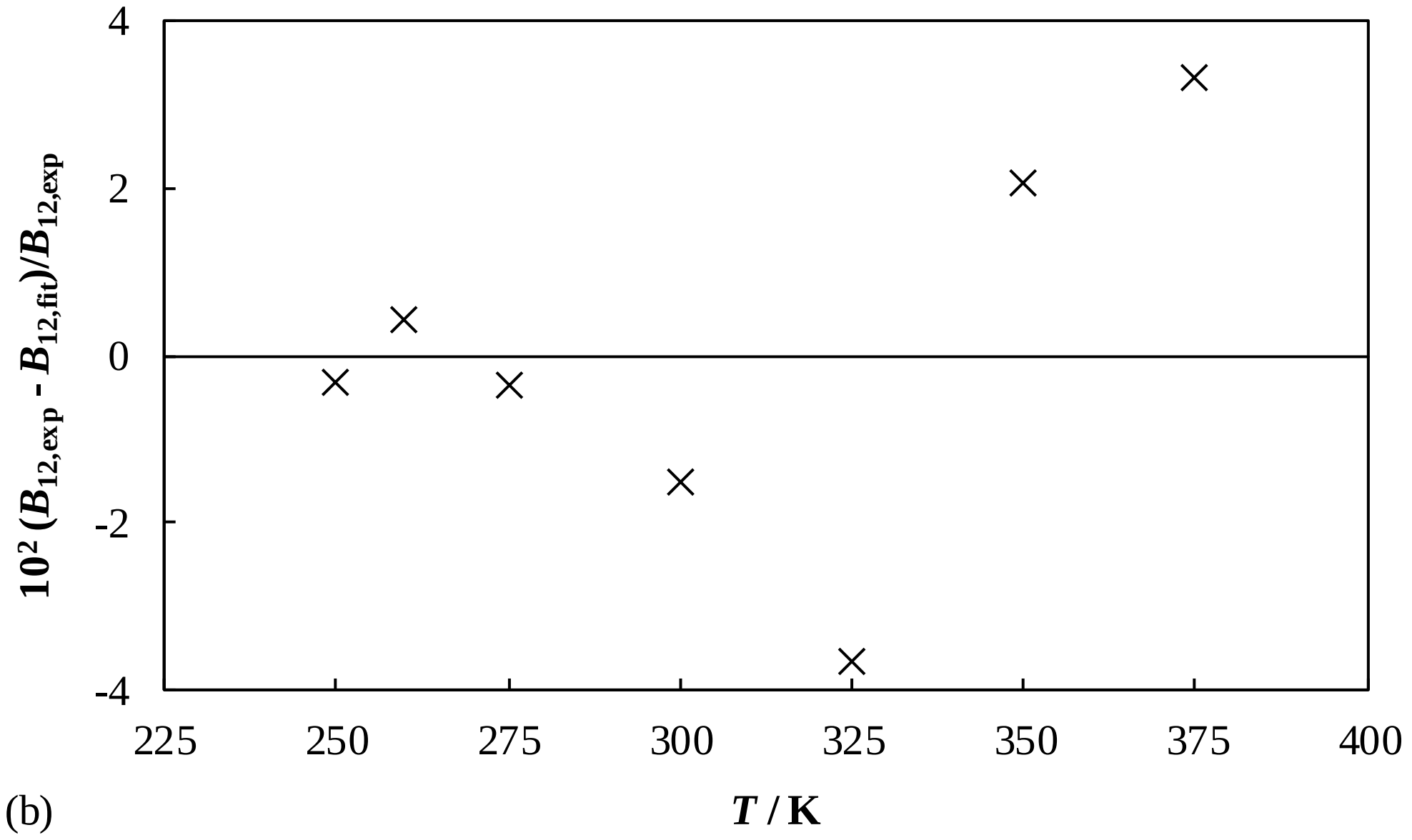

(b)

**Figure 6.** Relative deviations in (density or pressure) of the experimental ($p_{exp}$, $\rho_{exp}$, $T$) data of the five binary ($CO_2$ + $O_2$) mixtures measured in this work and in part one of this study [1], from (density or pressure) values calculated by the: (a) GERG-2008 [2], $\rho_{GERG}$, (b) EOS-CG [3], $\rho_{CG}$, and (c) virial equation of state, $p_{virial}$, as a function of density for different oxygen contents: ○ $x(O_2)$ = 0.05, × $x(O_2)$ = 0.10, △ $x(O_2)$ = 0.20, ◇ $x(O_2)$ = 0.50, and □ $x(O_2)$ = 0.75. Dashed lines indicate the expanded ($k$ = 2) uncertainty of the corresponding EoS (1 % threshold).

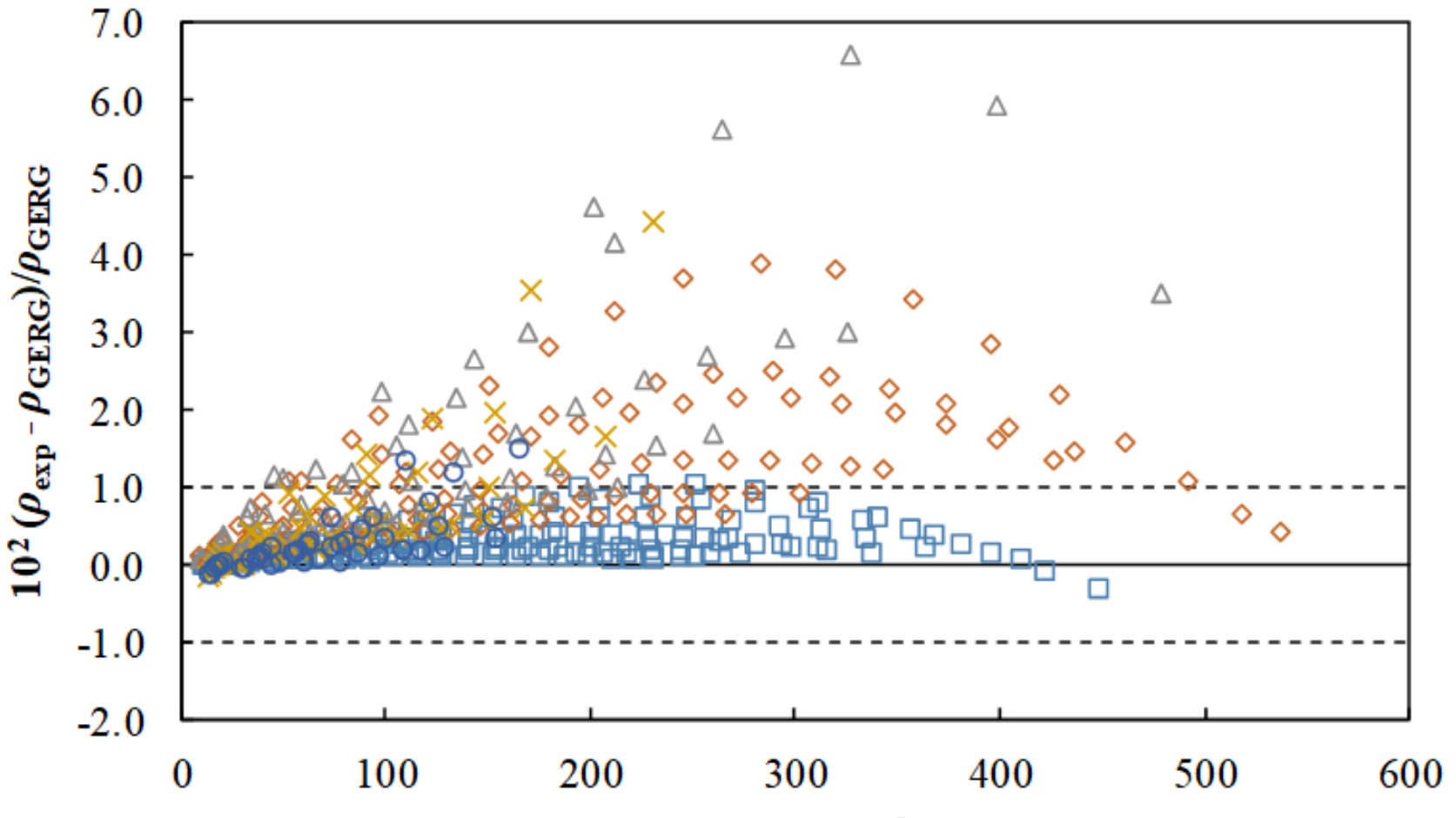

(a)

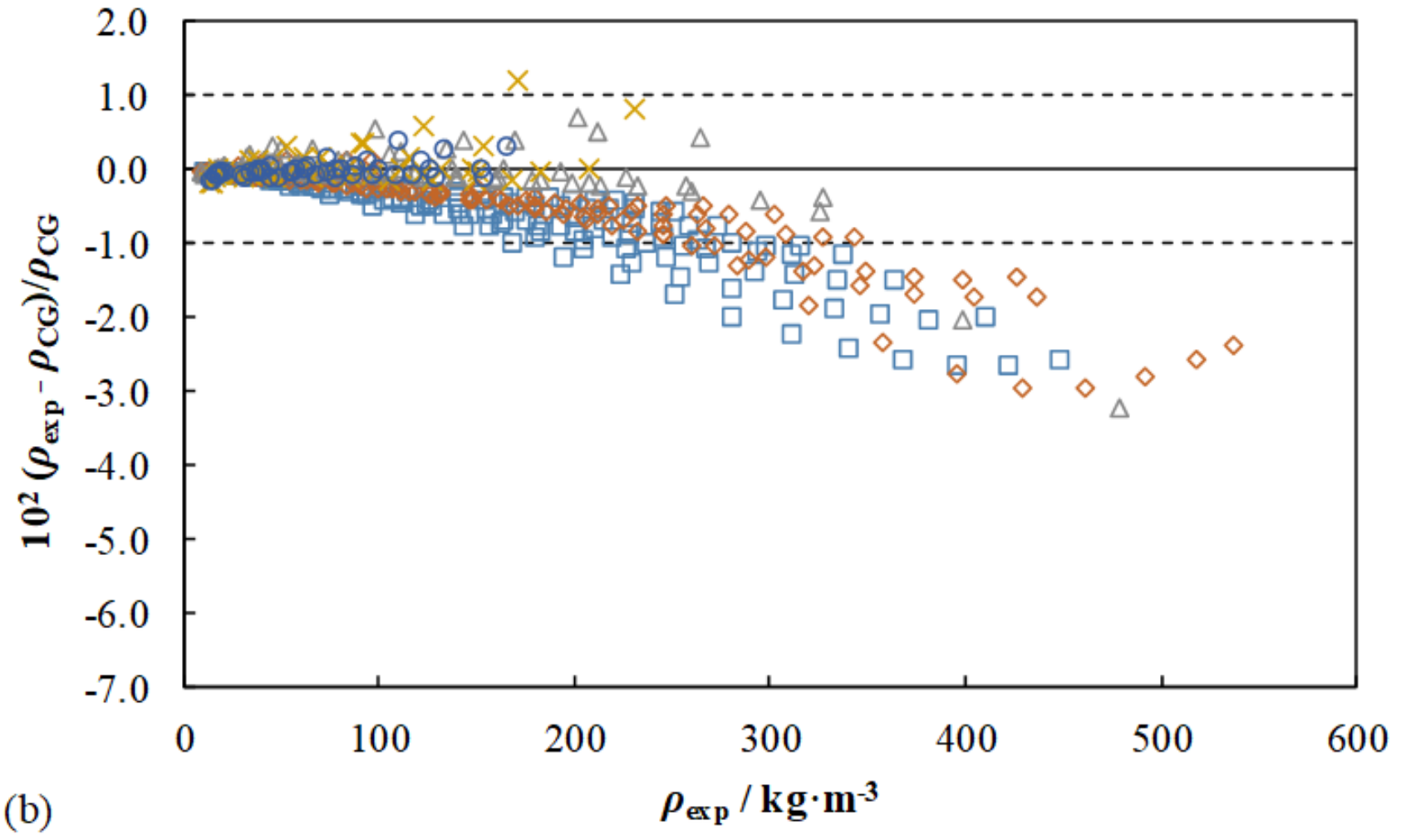

(b)

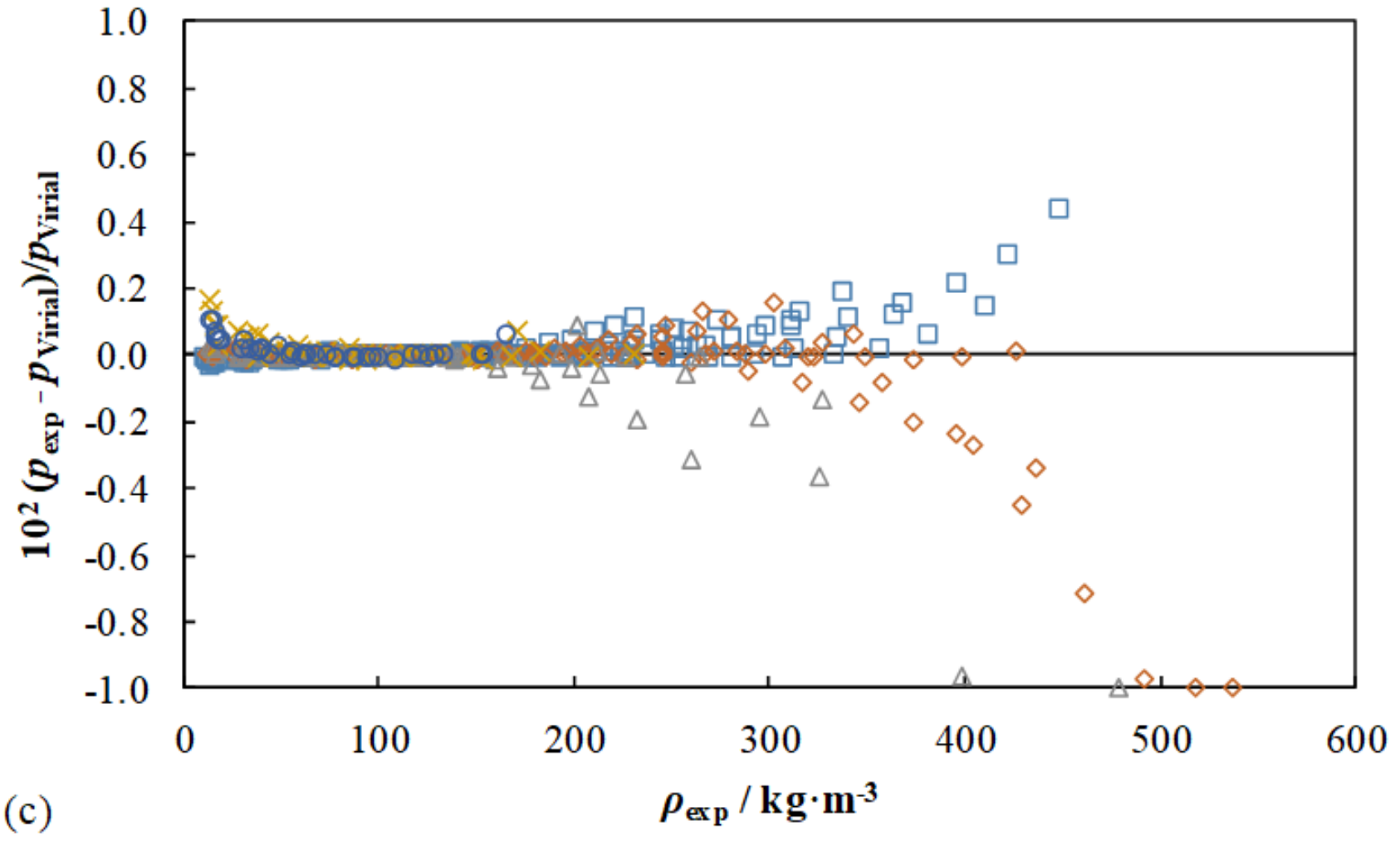

(c)

**Tables**

**Table 1.** Purity, supplier, molar mass, and critical parameters for the constituting components of the studied ($CO_2$ + $O_2$) mixtures in this work.

| | Purity / vol-% | Supplier | $M$ / g·mol$^{-1}$ | Critical parameters[a] | |
|---|---|---|---|---|---|
| | | | | $T_c$ / K | $p_c$ / MPa |
| Carbon dioxide | 99.9995 | Air Liquide | 44.010 | 304.13 | 7.3773 |
| Oxygen | 99.9999 | Linde | 31.999 | 154.58 | 5.0430 |

[a] Critical parameters were obtained by using the default equation for each substance in REFPROP software [29].

**Table 2.** Composition of the studied binary ($CO_2$ + $O_2$) mixtures in this work. Impurity compounds are marked in *italic* type.

| Component | (0.50 $CO_2$ + 0.50 $O_2$)[(a)] | | (0.25 $CO_2$ + 0.75 $O_2$)[(b)] | |
|---|---|---|---|---|
| | $10^2\ x_i$ / mol/mol | $10^2\ U(x_i)$ / mol/mol | $10^2\ x_i$ / mol/mol | $10^2\ U(x_i)$ / mol/mol |
| Carbon dioxide | 50.002657 | 0.000565 | 25.008606 | 0.000563 |
| Oxygen | 49.997192 | 0.000781 | 74.991234 | 0.000788 |
| *Argon* | 0.000050 | 0.000058 | 0.000075 | 0.000087 |
| *Nitrogen* | 0.000075 | 0.000065 | 0.000063 | 0.000052 |
| *Carbon monoxide* | 0.000020 | 0.000018 | 0.000015 | 0.000012 |
| *Propane* | 0.000004 | 0.000004 | 0.000005 | 0.000006 |
| *Nitric oxide* | 0.000002 | 0.000003 | 0.000001 | 0.000001 |
| Normalized composition without impurities | | | | |
| Carbon dioxide | 50.002733 | 0.000565 | 25.008646 | 0.000563 |
| Oxygen | 49.997267 | 0.000781 | 74.991354 | 0.000788 |

[(a)] BAM cylinder no.: 1009-180717

[(b)] BAM cylinder no.: 1099-180717

**Table 3.** Results of the gas chromatographic (GC) analysis and relative deviations between gravimetric preparation and GC analysis for the three ($CO_2$ + $O_2$) mixtures studied in this work. The results are followed by the gravimetric composition (non-normalized) of the employed validation mixtures.

| Component | Concentration | | Relative deviation between gravimetric composition and GC analysis |
|---|---|---|---|
| | $10^2\, x_i$ / mol/mol | $10^2\, U(x_i)$ / mol/mol | % |
| (0.50 $CO_2$ + 0.50 $O_2$) BAM cylinder no.: 1009-180717 | | | |
| Carbon dioxide | 49.9663 | 0.0539 | −0.073 |
| Oxygen | n. a. | n. a. | — |
| Validation mixture BAM cylinder no.: 96055001-980401 (G 050) | | | |
| Carbon dioxide | 49.906395 | 0.000918 | |
| Nitrogen | 50.093282 | 0.001397 | |
| *Oxygen* | 0.000240 | 0.000271 | |
| *Carbon monoxide* | 0.000077 | 0.000087 | |
| *Hydrogen* | 0.000006 | 0.000007 | |
| Validation mixture BAM cylinder no.: 8063-141006 (premixture G 473) | | | |
| Carbon dioxide | 51.479531 | 0.000953 | |
| Nitrogen | 44.109490 | 0.001903 | |
| Oxygen | 4.410942 | 0.000341 | |
| *Argon* | 0.000009 | 0.000010 | |
| *Carbon monoxide* | 0.000019 | 0.000018 | |
| *Methane* | 0.000001 | 0.000001 | |
| *Hydrogen* | 0.000007 | 0.000006 | |
| *Nitric oxide* | 0.000002 | 0.000003 | |
| (0.25 $CO_2$ + 0.75 $O_2$) BAM cylinder no.: 1099-180717 | | | |
| Carbon dioxide | 24.9755 | 0.0377 | −0.132 |
| Oxygen | n. a. | n. a. | — |
| Validation mixture BAM cylinder no.: C49312-010509 (G 050) | | | |
| Carbon dioxide | 25.020915 | 0.000536 | |
| Nitrogen | 74.979048 | 0.000842 | |
| *Oxygen* | 0.000016 | 0.000013 | |
| *Carbon monoxide* | 0.000011 | 0.000010 | |
| *Hydrogen* | 0.000009 | 0.000011 | |
| *Nitric oxide* | 0.000001 | 0.000001 | |
| Validation mixture BAM cylinder no.: 5018-020710 (G 050) | | | |
| Carbon dioxide | 27.264804 | 0.000521 | |
| Nitrogen | 72.735158 | 0.000819 | |
| *Oxygen* | 0.000016 | 0.000013 | |
| *Carbon monoxide* | 0.000012 | 0.000010 | |
| *Hydrogen* | 0.000009 | 0.000011 | |
| *Nitric oxide* | 0.000001 | 0.000002 | |

**Table 4.** Contributions to the expanded ($k$ = 2) overall uncertainty in density, $U_T(\rho_{exp})$, for the two ($CO_2$ + $O_2$) mixtures studied in this work.

| Source | Contribution ($k$ = 2) | Units | Estimation in density ($k$ = 2) | |
|---|---|---|---|---|
| | | | kg·m$^{-3}$ | % |
| | | ($0.50\ CO_2 + 0.50\ O_2$) | | |
| Temperature, $T$ | 0.004 | K | < 0.025 | < 0.0061 |
| Pressure, $p$ | < 0.005 | MPa | (0.045 – 0.18) | (0.021 – 0.73) |
| Composition, $x_i$ | < 0.0004 | mol·mol$^{-1}$ | < 0.010 | < 0.0022 |
| Density, $\rho$ | (0.036 – 0.099) | kg·m$^{-3}$ | (0.036 – 0.099) | (0.019 – 0.44) |
| Sum | | | (0.057 – 0.20) | (0.028 – 0.85) |
| | | ($0.25\ CO_2 + 0.75\ O_2$) | | |
| Temperature, $T$ | 0.004 | K | < 0.020 | < 0.0048 |
| Pressure, $p$ | < 0.005 | MPa | (0.041 – 0.14) | (0.025 – 0.37) |
| Composition, $x_i$ | < 0.0004 | mol·mol$^{-1}$ | < 0.073 | < 0.0017 |
| Density, $\rho$ | (0.042 – 0.101) | kg·m$^{-3}$ | (0.042 – 0.10) | (0.023 – 0.38) |
| Sum | | | (0.059 – 0.17) | (0.035 – 0.52) |

**Table 5.** Experimental ($p$, $\rho_{exp}$, $T$) measurements for the binary (0.50 $CO_2$ + 0.50 $O_2$) mixture, absolute and relative expanded ($k$ = 2) uncertainty in density, $U(\rho_{exp})$, and relative deviations from the density given by the GERG-2008 [2], $\rho_{GERG}$, and the EOS-CG [3], $\rho_{CG}$, equations of state.

| $T$ / K[(a)] | $p$ / MPa[(a)] | $\rho_{exp}$ / kg·m$^{-3}$ | $U(\rho_{exp})$ / kg·m$^{-3}$ | $10^2$ $U(\rho_{exp})/\rho_{exp}$ | $10^2$ ($\rho_{exp}$ - $\rho_{GERG})/\rho_{GERG}$ | $10^2$ ($\rho_{exp}$ - $\rho_{CG})/\rho_{CG}$ |
|---|---|---|---|---|---|---|
| 250 K isotherm | | | | | | |
| 250.046 | 2.551 | 52.505 | 0.046 | 0.087 | 1.10 | 0.12 |
| 250.047 | 2.006 | 40.148 | 0.044 | 0.110 | 0.81 | 0.08 |
| 250.044 | 1.420 | 27.631 | 0.043 | 0.155 | 0.53 | 0.05 |
| 250.045 | 0.994 | 18.956 | 0.042 | 0.221 | 0.34 | 0.02 |
| 250.044 | 0.498 | 9.294 | 0.041 | 0.438 | 0.12 | -0.03 |
| 260 K isotherm | | | | | | |
| 260.047 | 4.511 | 96.518 | 0.050 | 0.052 | 1.95 | 0.13 |
| 260.046 | 4.000 | 83.332 | 0.049 | 0.058 | 1.63 | 0.11 |
| 260.046 | 2.988 | 59.290 | 0.046 | 0.077 | 1.09 | 0.07 |
| 260.041 | 1.989 | 37.763 | 0.043 | 0.115 | 0.65 | 0.04 |
| 260.044 | 0.986 | 17.990 | 0.041 | 0.228 | 0.30 | 0.02 |
| 275 K isotherm | | | | | | |
| 275.018 | 18.783 | 536.868 | 0.099 | 0.019 | 0.44 | -2.36 |
| 275.016 | 18.027 | 518.230 | 0.097 | 0.019 | 0.68 | -2.57 |
| 275.014 | 17.023 | 491.321 | 0.094 | 0.019 | 1.08 | -2.81 |
| 275.012 | 16.026 | 461.967 | 0.091 | 0.020 | 1.59 | -2.95 |
| 275.012 | 15.013 | 429.453 | 0.087 | 0.020 | 2.22 | -2.94 |
| 275.010 | 14.020 | 395.120 | 0.083 | 0.021 | 2.87 | -2.74 |
| 275.008 | 13.018 | 358.480 | 0.079 | 0.022 | 3.45 | -2.34 |
| 275.008 | 12.009 | 320.447 | 0.075 | 0.023 | 3.82 | -1.83 |
| 275.007 | 11.009 | 282.804 | 0.070 | 0.025 | 3.90 | -1.31 |
| 275.005 | 10.006 | 246.161 | 0.066 | 0.027 | 3.69 | -0.89 |
| 275.003 | 9.005 | 211.687 | 0.062 | 0.029 | 3.29 | -0.59 |
| 275.004 | 8.005 | 179.703 | 0.059 | 0.033 | 2.81 | -0.39 |
| 275.003 | 7.002 | 150.278 | 0.055 | 0.037 | 2.32 | -0.26 |
| 275.003 | 6.001 | 123.367 | 0.052 | 0.042 | 1.85 | -0.18 |
| 275.003 | 5.000 | 98.668 | 0.049 | 0.050 | 1.43 | -0.13 |
| 275.002 | 3.999 | 75.936 | 0.047 | 0.062 | 1.06 | -0.09 |
| 275.001 | 3.004 | 55.017 | 0.044 | 0.081 | 0.74 | -0.06 |
| 275.001 | 1.999 | 35.366 | 0.042 | 0.119 | 0.45 | -0.04 |
| 275.002 | 0.999 | 17.108 | 0.040 | 0.234 | 0.21 | -0.02 |
| | | | 293.15 K isotherm | | | |
| 293.091 | 18.436 | 436.376 | 0.087 | 0.020 | 1.46 | -1.72 |
| 293.092 | 17.141 | 403.789 | 0.083 | 0.021 | 1.80 | -1.73 |
| 293.093 | 16.006 | 373.566 | 0.080 | 0.021 | 2.10 | -1.66 |
| 293.093 | 15.005 | 345.809 | 0.077 | 0.022 | 2.30 | -1.55 |
| 293.094 | 14.010 | 317.600 | 0.073 | 0.023 | 2.44 | -1.39 |
| 293.093 | 13.006 | 288.886 | 0.070 | 0.024 | 2.50 | -1.20 |
| 293.094 | 12.007 | 260.508 | 0.067 | 0.026 | 2.46 | -1.01 |
| 293.094 | 11.007 | 232.662 | 0.064 | 0.027 | 2.35 | -0.82 |
| 293.092 | 10.043 | 206.648 | 0.061 | 0.029 | 2.17 | -0.67 |
| 293.092 | 9.011 | 179.946 | 0.058 | 0.032 | 1.94 | -0.53 |
| 293.094 | 8.003 | 155.119 | 0.055 | 0.035 | 1.70 | -0.41 |
| 293.092 | 7.001 | 131.760 | 0.052 | 0.040 | 1.46 | -0.31 |
| 293.092 | 6.009 | 109.848 | 0.050 | 0.045 | 1.21 | -0.24 |
| 293.092 | 5.000 | 88.810 | 0.047 | 0.053 | 0.96 | -0.18 |
| 293.093 | 3.999 | 69.090 | 0.045 | 0.065 | 0.74 | -0.13 |
| 293.092 | 2.985 | 50.178 | 0.043 | 0.086 | 0.52 | -0.09 |
| 293.092 | 1.999 | 32.751 | 0.041 | 0.125 | 0.34 | -0.05 |
| 293.092 | 0.999 | 15.954 | 0.039 | 0.245 | 0.16 | -0.02 |
| | | | 300 K isotherm | | | |
| 299.947 | 19.218 | 426.654 | 0.086 | 0.020 | 1.38 | -1.44 |
| 299.946 | 18.017 | 398.813 | 0.082 | 0.021 | 1.62 | -1.47 |
| 299.947 | 17.009 | 374.312 | 0.080 | 0.021 | 1.81 | -1.44 |
| 299.946 | 16.013 | 349.239 | 0.077 | 0.022 | 1.97 | -1.38 |
| 299.944 | 15.010 | 323.405 | 0.074 | 0.023 | 2.09 | -1.28 |
| 299.947 | 14.010 | 297.246 | 0.071 | 0.024 | 2.16 | -1.16 |
| 299.945 | 13.010 | 271.079 | 0.068 | 0.025 | 2.16 | -1.02 |
| 299.947 | 12.006 | 245.004 | 0.065 | 0.026 | 2.10 | -0.88 |
| 299.945 | 11.006 | 219.591 | 0.062 | 0.028 | 1.99 | -0.74 |
| 299.946 | 10.005 | 194.857 | 0.059 | 0.030 | 1.83 | -0.61 |
| 299.945 | 9.004 | 171.008 | 0.056 | 0.033 | 1.65 | -0.50 |
| 299.947 | 8.002 | 148.143 | 0.054 | 0.036 | 1.45 | -0.41 |
| 299.948 | 7.002 | 126.328 | 0.051 | 0.041 | 1.25 | -0.32 |
| 299.946 | 6.038 | 106.284 | 0.049 | 0.046 | 1.05 | -0.25 |
| 299.947 | 5.008 | 85.884 | 0.047 | 0.054 | 0.84 | -0.19 |
| 299.945 | 4.001 | 66.926 | 0.045 | 0.067 | 0.65 | -0.13 |
| 299.945 | 3.000 | 48.979 | 0.042 | 0.087 | 0.47 | -0.09 |
| 299.946 | 1.999 | 31.881 | 0.041 | 0.127 | 0.31 | -0.05 |
| 299.946 | 0.999 | 15.565 | 0.039 | 0.249 | 0.15 | -0.03 |
| | | | 325 K isotherm | | | |
| 324.952 | 18.906 | 343.259 | 0.075 | 0.022 | 1.25 | -0.91 |
| 324.952 | 18.076 | 327.073 | 0.073 | 0.022 | 1.30 | -0.90 |
| 324.953 | 17.106 | 307.850 | 0.071 | 0.023 | 1.34 | -0.89 |
| 324.952 | 16.105 | 287.748 | 0.069 | 0.024 | 1.36 | -0.85 |
| 324.952 | 15.122 | 267.889 | 0.066 | 0.025 | 1.37 | -0.81 |
| 324.953 | 14.008 | 245.322 | 0.064 | 0.026 | 1.35 | -0.75 |
| 324.952 | 13.013 | 225.252 | 0.062 | 0.027 | 1.31 | -0.69 |
| 324.953 | 11.990 | 204.805 | 0.059 | 0.029 | 1.25 | -0.63 |
| 324.953 | 11.012 | 185.551 | 0.057 | 0.031 | 1.18 | -0.55 |
| 324.953 | 10.011 | 166.210 | 0.055 | 0.033 | 1.09 | -0.49 |
| 324.952 | 8.969 | 146.502 | 0.053 | 0.036 | 0.98 | -0.42 |
| 324.953 | 8.001 | 128.664 | 0.051 | 0.039 | 0.88 | -0.36 |
| 324.953 | 7.004 | 110.806 | 0.048 | 0.044 | 0.77 | -0.29 |
| 324.953 | 6.002 | 93.365 | 0.046 | 0.050 | 0.65 | -0.24 |
| 324.953 | 5.001 | 76.501 | 0.045 | 0.058 | 0.54 | -0.19 |
| 324.951 | 3.999 | 60.141 | 0.043 | 0.071 | 0.42 | -0.15 |
| 324.951 | 2.986 | 44.141 | 0.041 | 0.093 | 0.31 | -0.11 |
| 324.951 | 2.000 | 29.078 | 0.039 | 0.135 | 0.20 | -0.07 |
| 324.951 | 0.999 | 14.286 | 0.037 | 0.262 | 0.09 | -0.04 |
| | | | 350 K isotherm | | | |
| 349.935 | 19.393 | 302.011 | 0.069 | 0.023 | 0.93 | -0.62 |
| 349.936 | 18.003 | 279.264 | 0.067 | 0.024 | 0.95 | -0.62 |
| 349.935 | 17.004 | 262.723 | 0.065 | 0.025 | 0.95 | -0.61 |
| 349.935 | 16.001 | 246.017 | 0.063 | 0.026 | 0.95 | -0.59 |
| 349.935 | 15.004 | 229.333 | 0.061 | 0.027 | 0.93 | -0.57 |
| 349.934 | 14.003 | 212.603 | 0.059 | 0.028 | 0.91 | -0.54 |
| 349.934 | 13.000 | 195.872 | 0.057 | 0.029 | 0.87 | -0.51 |
| 349.935 | 12.001 | 179.322 | 0.055 | 0.031 | 0.83 | -0.47 |
| 349.935 | 11.001 | 162.916 | 0.054 | 0.033 | 0.78 | -0.42 |
| 349.934 | 10.021 | 147.007 | 0.052 | 0.035 | 0.72 | -0.38 |
| 349.936 | 9.000 | 130.648 | 0.050 | 0.038 | 0.66 | -0.34 |
| 349.935 | 8.000 | 114.907 | 0.048 | 0.042 | 0.59 | -0.29 |
| 349.935 | 7.000 | 99.435 | 0.046 | 0.047 | 0.52 | -0.25 |
| 349.935 | 6.000 | 84.254 | 0.045 | 0.053 | 0.45 | -0.21 |
| 349.935 | 5.000 | 69.392 | 0.043 | 0.062 | 0.37 | -0.17 |
| 349.936 | 3.999 | 54.846 | 0.041 | 0.075 | 0.29 | -0.13 |
| 349.936 | 2.985 | 40.432 | 0.040 | 0.098 | 0.20 | -0.11 |
| 349.935 | 1.999 | 26.749 | 0.038 | 0.142 | 0.12 | -0.08 |
| 349.935 | 1.000 | 13.220 | 0.036 | 0.276 | 0.05 | -0.05 |
| | | | 375 K isotherm | | | |
| 374.922 | 19.304 | 265.337 | 0.064 | 0.024 | 0.67 | -0.48 |
| 374.922 | 18.002 | 246.901 | 0.062 | 0.025 | 0.67 | -0.48 |
| 374.922 | 16.998 | 232.559 | 0.061 | 0.026 | 0.67 | -0.48 |
| 374.921 | 15.996 | 218.186 | 0.059 | 0.027 | 0.66 | -0.47 |
| 374.920 | 14.998 | 203.822 | 0.057 | 0.028 | 0.64 | -0.45 |
| 374.921 | 14.017 | 189.691 | 0.056 | 0.029 | 0.62 | -0.43 |
| 374.921 | 12.998 | 175.041 | 0.054 | 0.031 | 0.59 | -0.41 |
| 374.920 | 11.999 | 160.727 | 0.052 | 0.033 | 0.56 | -0.38 |
| 374.920 | 10.999 | 146.474 | 0.051 | 0.035 | 0.53 | -0.35 |
| 374.919 | 10.000 | 132.318 | 0.049 | 0.037 | 0.49 | -0.32 |
| 374.921 | 9.000 | 118.279 | 0.048 | 0.040 | 0.45 | -0.29 |
| 374.920 | 7.999 | 104.368 | 0.046 | 0.044 | 0.40 | -0.25 |
| 374.921 | 6.999 | 90.633 | 0.044 | 0.049 | 0.36 | -0.21 |
| 374.922 | 6.000 | 77.083 | 0.043 | 0.056 | 0.31 | -0.18 |
| 374.920 | 5.000 | 63.711 | 0.041 | 0.065 | 0.25 | -0.15 |
| 374.920 | 4.005 | 50.603 | 0.040 | 0.079 | 0.20 | -0.12 |
| 374.921 | 2.998 | 37.555 | 0.038 | 0.102 | 0.15 | -0.10 |
| 374.922 | 2.000 | 24.835 | 0.037 | 0.149 | 0.10 | -0.06 |
| 374.921 | 0.999 | 12.288 | 0.036 | 0.289 | 0.05 | -0.03 |

(a) Expanded uncertainties ($k$ = 2): $U(p > 3)/\mathrm{MPa} = 75 \cdot 10^{-6} \cdot \frac{p}{\mathrm{MPa}} + 3.5 \cdot 10^{-3}$; $U(p < 3)/\mathrm{MPa} = 60 \cdot 10^{-6} \cdot \frac{p}{\mathrm{MPa}} + 1.7 \cdot 10^{-3}$; $U(T)$ = 4 mK; $\frac{U(\rho)}{\mathrm{kg \cdot m^{-3}}} = 2.5 \cdot 10^{4} \frac{\chi s}{\mathrm{m^{3} \cdot kg^{-1}}} + 1.1 \cdot 10^{-4} \cdot \frac{\rho}{\mathrm{kg \cdot m^{-3}}} + 2.3 \cdot 10^{-2}$.

**Table 6.** Experimental ($p$, $\rho_{exp}$, $T$) measurements for the binary (0.25 $CO_2$ + 0.75 $O_2$) mixture, absolute and relative expanded ($k$ = 2) uncertainty in density, $U(\rho_{exp})$, and relative deviations from the density given by the GERG-2008 [2], $\rho_{GERG}$, and the EOS-CG [3], $\rho_{CG}$, equations of state.

| $T$ / K[(a)] | $p$ / MPa[(a)] | $\rho_{exp}$ / kg·m$^{-3}$[(a)] | $U(\rho_{exp})$ / kg·m$^{-3}$ | $10^2$ $U(\rho_{exp})/\rho_{exp}$ | $10^2$ ($\rho_{exp}$ - $\rho_{GERG}$)/$\rho_{GERG}$ | $10^2$ ($\rho_{exp}$ - $\rho_{CG}$)/$\rho_{CG}$ |
|---|---|---|---|---|---|---|
| | | | 250 K isotherm | | | |
| 250.033 | 18.094 | 448.577 | 0.101 | 0.023 | -0.28 | -2.57 |
| 250.034 | 17.025 | 422.254 | 0.098 | 0.023 | -0.06 | -2.64 |
| 250.031 | 16.013 | 395.828 | 0.095 | 0.024 | 0.17 | -2.64 |
| 250.034 | 15.018 | 368.491 | 0.092 | 0.025 | 0.41 | -2.56 |
| 250.035 | 14.016 | 339.912 | 0.089 | 0.026 | 0.63 | -2.42 |
| 250.035 | 13.015 | 310.671 | 0.086 | 0.028 | 0.82 | -2.22 |
| 250.034 | 12.010 | 281.022 | 0.082 | 0.029 | 0.96 | -1.97 |
| 250.032 | 11.005 | 251.582 | 0.079 | 0.031 | 1.04 | -1.69 |
| 250.033 | 10.007 | 222.915 | 0.076 | 0.034 | 1.05 | -1.42 |
| 250.032 | 9.005 | 195.078 | 0.072 | 0.037 | 1.00 | -1.17 |
| 250.021 | 8.017 | 168.747 | 0.069 | 0.041 | 0.90 | -0.97 |
| 250.015 | 7.001 | 143.020 | 0.066 | 0.046 | 0.80 | -0.77 |
| 250.012 | 6.002 | 119.016 | 0.064 | 0.054 | 0.67 | -0.61 |
| 250.013 | 5.000 | 96.291 | 0.061 | 0.063 | 0.54 | -0.47 |
| 250.012 | 3.999 | 74.849 | 0.059 | 0.078 | 0.41 | -0.35 |
| 250.021 | 2.975 | 54.134 | 0.056 | 0.104 | 0.32 | -0.21 |
| 250.031 | 1.999 | 35.419 | 0.054 | 0.153 | 0.21 | -0.12 |
| 250.029 | 0.999 | 17.245 | 0.052 | 0.302 | 0.11 | -0.05 |
| | | | 260 K isotherm | | | |
| 260.029 | 18.272 | 410.337 | 0.096 | 0.023 | 0.10 | -2.00 |
| 260.030 | 17.019 | 381.362 | 0.093 | 0.024 | 0.30 | -2.01 |
| 260.029 | 16.011 | 357.028 | 0.090 | 0.025 | 0.46 | -1.96 |
| 260.028 | 15.031 | 332.575 | 0.087 | 0.026 | 0.60 | -1.88 |
| 260.028 | 14.021 | 306.829 | 0.084 | 0.027 | 0.73 | -1.76 |
| 260.026 | 13.011 | 280.775 | 0.081 | 0.029 | 0.82 | -1.61 |
| 260.025 | 12.009 | 254.904 | 0.078 | 0.031 | 0.88 | -1.44 |
| 260.027 | 11.006 | 229.285 | 0.075 | 0.033 | 0.91 | -1.25 |
| 260.027 | 10.003 | 204.098 | 0.072 | 0.035 | 0.88 | -1.07 |
| 260.025 | 9.003 | 179.682 | 0.070 | 0.039 | 0.84 | -0.90 |
| 260.027 | 8.005 | 156.109 | 0.067 | 0.043 | 0.76 | -0.75 |
| 260.026 | 7.006 | 133.454 | 0.064 | 0.048 | 0.68 | -0.60 |
| 260.025 | 6.005 | 111.681 | 0.062 | 0.055 | 0.59 | -0.46 |
| 260.025 | 5.013 | 91.047 | 0.059 | 0.065 | 0.50 | -0.35 |
| 260.026 | 4.001 | 70.943 | 0.057 | 0.081 | 0.40 | -0.25 |
| 260.023 | 3.000 | 51.968 | 0.055 | 0.106 | 0.30 | -0.16 |
| 260.024 | 2.000 | 33.855 | 0.053 | 0.156 | 0.21 | -0.08 |
| 260.024 | 0.999 | 16.536 | 0.051 | 0.308 | 0.11 | -0.02 |
| | | | 275 K isotherm | | | |
| 274.994 | 18.413 | 363.002 | 0.089 | 0.025 | 0.25 | -1.50 |
| 274.992 | 17.016 | 334.214 | 0.086 | 0.026 | 0.38 | -1.47 |
| 274.996 | 16.006 | 312.797 | 0.083 | 0.027 | 0.46 | -1.42 |
| 274.998 | 15.022 | 291.571 | 0.081 | 0.028 | 0.53 | -1.36 |
| 274.997 | 14.008 | 269.483 | 0.078 | 0.029 | 0.59 | -1.27 |
| 274.998 | 13.009 | 247.616 | 0.076 | 0.031 | 0.63 | -1.17 |
| 274.998 | 12.008 | 225.785 | 0.073 | 0.033 | 0.64 | -1.07 |
| 274.998 | 11.003 | 204.069 | 0.071 | 0.035 | 0.65 | -0.95 |
| 274.998 | 10.004 | 182.777 | 0.068 | 0.037 | 0.62 | -0.83 |
| 274.998 | 9.002 | 161.854 | 0.066 | 0.041 | 0.59 | -0.72 |
| 274.997 | 8.002 | 141.471 | 0.064 | 0.045 | 0.54 | -0.61 |
| 274.997 | 7.010 | 121.823 | 0.062 | 0.051 | 0.49 | -0.50 |
| 274.996 | 6.013 | 102.658 | 0.059 | 0.058 | 0.43 | -0.40 |
| 274.996 | 5.032 | 84.400 | 0.057 | 0.068 | 0.37 | -0.31 |
| 274.996 | 4.005 | 65.923 | 0.055 | 0.084 | 0.30 | -0.22 |
| 274.995 | 2.995 | 48.398 | 0.053 | 0.110 | 0.23 | -0.14 |
| 274.995 | 1.999 | 31.732 | 0.051 | 0.162 | 0.16 | -0.07 |
| 274.996 | 0.999 | 15.571 | 0.049 | 0.317 | 0.10 | -0.01 |
| 293.15 K isotherm | | | | | | |
| 293.086 | 19.488 | 337.017 | 0.085 | 0.025 | 0.17 | -1.15 |
| 293.082 | 18.016 | 311.066 | 0.082 | 0.026 | 0.25 | -1.14 |
| 293.079 | 17.009 | 292.913 | 0.080 | 0.027 | 0.30 | -1.12 |
| 293.079 | 15.615 | 267.384 | 0.077 | 0.029 | 0.36 | -1.06 |
| 293.078 | 15.008 | 256.154 | 0.075 | 0.029 | 0.38 | -1.03 |
| 293.082 | 13.998 | 237.416 | 0.073 | 0.031 | 0.41 | -0.97 |
| 293.085 | 13.004 | 218.912 | 0.071 | 0.032 | 0.42 | -0.90 |
| 293.088 | 12.006 | 200.401 | 0.069 | 0.034 | 0.43 | -0.83 |
| 293.088 | 11.004 | 181.957 | 0.067 | 0.037 | 0.43 | -0.75 |
| 293.088 | 10.004 | 163.705 | 0.065 | 0.040 | 0.41 | -0.67 |
| 293.089 | 9.002 | 145.672 | 0.063 | 0.043 | 0.39 | -0.59 |
| 293.089 | 8.002 | 127.961 | 0.061 | 0.047 | 0.36 | -0.50 |
| 293.089 | 7.001 | 110.579 | 0.059 | 0.053 | 0.33 | -0.42 |
| 293.090 | 6.001 | 93.576 | 0.057 | 0.061 | 0.29 | -0.34 |
| 293.088 | 5.000 | 76.953 | 0.055 | 0.071 | 0.24 | -0.27 |
| 293.088 | 3.999 | 60.720 | 0.053 | 0.087 | 0.20 | -0.20 |
| 293.089 | 2.992 | 44.801 | 0.051 | 0.114 | 0.15 | -0.13 |
| 293.089 | 2.001 | 29.567 | 0.050 | 0.167 | 0.11 | -0.07 |
| 293.090 | 1.002 | 14.596 | 0.048 | 0.328 | 0.07 | -0.02 |
| | | | 300 K isotherm | | | |
| 299.942 | 19.099 | 316.114 | 0.082 | 0.026 | 0.21 | -1.02 |
| 299.942 | 18.014 | 297.775 | 0.080 | 0.027 | 0.26 | -1.01 |
| 299.942 | 17.010 | 280.533 | 0.078 | 0.028 | 0.30 | -0.99 |
| 299.944 | 16.011 | 263.168 | 0.076 | 0.029 | 0.33 | -0.96 |
| 299.942 | 15.009 | 245.604 | 0.074 | 0.030 | 0.36 | -0.92 |
| 299.943 | 14.009 | 227.985 | 0.072 | 0.031 | 0.38 | -0.87 |
| 299.943 | 13.007 | 210.309 | 0.070 | 0.033 | 0.39 | -0.81 |
| 299.944 | 12.005 | 192.673 | 0.068 | 0.035 | 0.39 | -0.75 |
| 299.944 | 11.004 | 175.153 | 0.066 | 0.037 | 0.39 | -0.67 |
| 299.942 | 10.004 | 157.797 | 0.064 | 0.040 | 0.37 | -0.60 |
| 299.942 | 9.002 | 140.615 | 0.062 | 0.044 | 0.35 | -0.53 |
| 299.941 | 8.013 | 123.873 | 0.060 | 0.048 | 0.33 | -0.46 |
| 299.942 | 7.006 | 107.131 | 0.058 | 0.054 | 0.30 | -0.38 |
| 299.941 | 6.023 | 91.070 | 0.056 | 0.061 | 0.26 | -0.31 |
| 299.943 | 5.002 | 74.728 | 0.054 | 0.072 | 0.22 | -0.24 |
| 299.942 | 4.001 | 59.065 | 0.052 | 0.089 | 0.18 | -0.18 |
| 299.942 | 2.963 | 43.181 | 0.051 | 0.117 | 0.14 | -0.12 |
| 299.941 | 1.999 | 28.786 | 0.049 | 0.170 | 0.11 | -0.06 |
| 299.944 | 0.999 | 14.204 | 0.047 | 0.333 | 0.07 | -0.01 |
| | | | 325 K isotherm | | | |
| 324.952 | 18.940 | 272.696 | 0.075 | 0.027 | 0.16 | -0.75 |
| 324.952 | 18.004 | 259.112 | 0.073 | 0.028 | 0.18 | -0.74 |
| 324.952 | 17.018 | 244.643 | 0.072 | 0.029 | 0.20 | -0.73 |
| 324.952 | 16.007 | 229.703 | 0.070 | 0.031 | 0.22 | -0.70 |
| 324.953 | 15.007 | 214.817 | 0.068 | 0.032 | 0.23 | -0.67 |
| 324.952 | 14.007 | 199.880 | 0.067 | 0.033 | 0.24 | -0.64 |
| 324.952 | 13.004 | 184.876 | 0.065 | 0.035 | 0.25 | -0.60 |
| 324.952 | 12.002 | 169.876 | 0.063 | 0.037 | 0.25 | -0.56 |
| 324.952 | 11.002 | 154.959 | 0.062 | 0.040 | 0.25 | -0.51 |
| 324.951 | 10.002 | 140.094 | 0.060 | 0.043 | 0.24 | -0.46 |
| 324.951 | 9.002 | 125.321 | 0.058 | 0.046 | 0.22 | -0.41 |
| 324.951 | 8.001 | 110.662 | 0.057 | 0.051 | 0.21 | -0.36 |
| 324.952 | 7.008 | 96.262 | 0.055 | 0.057 | 0.19 | -0.30 |
| 324.951 | 6.005 | 81.875 | 0.053 | 0.065 | 0.17 | -0.25 |
| 324.953 | 4.998 | 67.607 | 0.052 | 0.076 | 0.14 | -0.20 |
| 324.952 | 3.993 | 53.574 | 0.050 | 0.093 | 0.11 | -0.15 |
| 324.952 | 2.991 | 39.794 | 0.048 | 0.122 | 0.09 | -0.10 |
| 324.952 | 2.001 | 26.394 | 0.047 | 0.178 | 0.07 | -0.05 |
| 324.954 | 0.995 | 13.010 | 0.045 | 0.349 | 0.04 | -0.02 |
| | | | 350 K isotherm | | | |
| 349.941 | 19.610 | 251.685 | 0.071 | 0.028 | 0.12 | -0.55 |
| 349.940 | 19.007 | 244.049 | 0.070 | 0.029 | 0.13 | -0.55 |
| 349.939 | 18.006 | 231.282 | 0.069 | 0.030 | 0.15 | -0.54 |
| 349.938 | 17.002 | 218.364 | 0.067 | 0.031 | 0.16 | -0.53 |
| 349.937 | 16.003 | 205.413 | 0.066 | 0.032 | 0.17 | -0.51 |
| 349.936 | 15.002 | 192.363 | 0.064 | 0.033 | 0.18 | -0.49 |
| 349.936 | 14.004 | 179.307 | 0.063 | 0.035 | 0.19 | -0.47 |
| 349.937 | 13.001 | 166.136 | 0.061 | 0.037 | 0.19 | -0.44 |
| 349.938 | 12.001 | 152.987 | 0.060 | 0.039 | 0.19 | -0.41 |
| 349.937 | 11.002 | 139.860 | 0.058 | 0.042 | 0.19 | -0.37 |
| 349.937 | 10.001 | 126.717 | 0.057 | 0.045 | 0.18 | -0.34 |
| 349.938 | 9.000 | 113.615 | 0.055 | 0.049 | 0.17 | -0.30 |
| 349.938 | 8.000 | 100.577 | 0.054 | 0.054 | 0.16 | -0.26 |
| 349.936 | 7.002 | 87.645 | 0.052 | 0.060 | 0.15 | -0.22 |
| 349.937 | 6.010 | 74.879 | 0.051 | 0.068 | 0.13 | -0.18 |
| 349.938 | 5.000 | 61.978 | 0.049 | 0.080 | 0.11 | -0.14 |
| 349.937 | 4.031 | 49.701 | 0.048 | 0.097 | 0.09 | -0.11 |
| 349.937 | 2.989 | 36.636 | 0.047 | 0.127 | 0.08 | -0.07 |
| 349.937 | 1.999 | 24.360 | 0.045 | 0.185 | 0.06 | -0.03 |
| 349.938 | 0.999 | 12.096 | 0.044 | 0.362 | 0.03 | -0.01 |
| | | | 375 K isotherm | | | |
| 374.924 | 19.829 | 230.614 | 0.067 | 0.029 | 0.08 | -0.43 |
| 374.925 | 19.006 | 221.282 | 0.066 | 0.030 | 0.09 | -0.42 |
| 374.924 | 17.999 | 209.763 | 0.065 | 0.031 | 0.11 | -0.42 |
| 374.922 | 17.000 | 198.252 | 0.064 | 0.032 | 0.12 | -0.41 |
| 374.923 | 16.002 | 186.676 | 0.062 | 0.033 | 0.12 | -0.39 |
| 374.922 | 14.999 | 174.971 | 0.061 | 0.035 | 0.13 | -0.38 |
| 374.922 | 13.998 | 163.244 | 0.060 | 0.037 | 0.14 | -0.36 |
| 374.922 | 12.999 | 151.487 | 0.058 | 0.039 | 0.14 | -0.34 |
| 374.923 | 11.978 | 139.434 | 0.057 | 0.041 | 0.14 | -0.31 |
| 374.922 | 11.000 | 127.886 | 0.056 | 0.044 | 0.14 | -0.28 |
| 374.923 | 10.000 | 116.053 | 0.054 | 0.047 | 0.13 | -0.26 |
| 374.922 | 9.000 | 104.240 | 0.053 | 0.051 | 0.12 | -0.23 |
| 374.922 | 8.000 | 92.436 | 0.052 | 0.056 | 0.11 | -0.20 |
| 374.922 | 6.999 | 80.668 | 0.050 | 0.062 | 0.10 | -0.17 |
| 374.923 | 5.999 | 68.937 | 0.049 | 0.071 | 0.09 | -0.14 |
| 374.924 | 4.999 | 57.260 | 0.048 | 0.083 | 0.08 | -0.12 |
| 374.923 | 3.999 | 45.645 | 0.046 | 0.102 | 0.06 | -0.09 |
| 374.924 | 2.984 | 33.935 | 0.045 | 0.133 | 0.05 | -0.06 |
| 374.925 | 1.999 | 22.640 | 0.044 | 0.193 | 0.03 | -0.04 |
| 374.923 | 0.998 | 11.262 | 0.042 | 0.377 | 0.02 | -0.02 |

[(a)] Expanded uncertainties ($k$ = 2): $U(p > 3)/\mathrm{MPa} = 75 \cdot 10^{-6} \cdot \frac{p}{\mathrm{MPa}} + 3.5 \cdot 10^{-3}$; $U(p < 3)/\mathrm{MPa} = 60 \cdot 10^{-6} \cdot \frac{p}{\mathrm{MPa}} + 1.7 \cdot 10^{-3}$; $U(T)$ = 4 mK; $\frac{U(\rho)}{\mathrm{kg \cdot m^{-3}}} = 2.5 \cdot 10^{4} \frac{\chi_\mathrm{S}}{\mathrm{m^3 \cdot kg^{-1}}} + 1.1 \cdot 10^{-4} \cdot \frac{\rho}{\mathrm{kg \cdot m^{-3}}} + 2.3 \cdot 10^{-2}$.

**Table 7.** Virial coefficients $B(T)$ and $C(T)$ and second interaction virial coefficient $B_{12}(T)$, with their expanded ($k$ = 2) uncertainties, from the fit to the five experimental binary ($CO_2$ + $O_2$) mixtures studied in this work and part one of this study [1], at the average temperature of each isotherm.

| $T$ / K | $B$ / $cm^3 \cdot mol^{-1}$ | $U(B)$ / $cm^3 \cdot mol^{-1}$ | $C$ / $cm^6 \cdot mol^{-2}$ | $U(C)$ / $cm^6 \cdot mol^{-2}$ | $B_{12}$ / $cm^3 \cdot mol^{-1}$ | $U(B_{12})$ / $cm^3 \cdot mol^{-1}$ | $10^2 (B_{12,exp} - B_{12,GERG})/B_{12,GERG}$ | $10^2 (B_{12,exp} - B_{12,CG})/B_{12,CG}$ |
|---|---|---|---|---|---|---|---|---|
| | | | | ($0.95\ CO_2 + 0.05\ O_2$) | | | | |
| 299.947 | -113.22 | 0.94 | 4474 | 365 | -39.39 | 1.97 | 68.2 | 2.7 |
| 324.953 | -93.82 | 0.28 | 3943 | 91 | -30.83 | 0.81 | 59.2 | -3.1 |
| | | | | ($0.90\ CO_2 + 0.10\ O_2$) | | | | |
| 299.924 | -105.90 | 0.20 | 4294 | 40 | -41.31 | 0.71 | 66.6 | 6.3 |
| 324.937 | -87.07 | 0.19 | 3618 | 43 | -29.65 | 0.69 | 45.1 | -7.8 |
| 374.913 | -61.23 | 0.32 | 2866 | 105 | -19.37 | 0.87 | 53.2 | -8.0 |
| | | | | ($0.80\ CO_2 + 0.20\ O_2$) | | | | |
| 299.947 | -91.56 | 0.12 | 3746 | 21 | -41.52 | 0.63 | 51.6 | 4.3 |
| 324.955 | -75.35 | 0.20 | 3238 | 49 | -32.65 | 0.71 | 45.6 | -0.4 |
| 349.939 | -62.69 | 0.62 | 2922 | 245 | -26.14 | 1.38 | 45.7 | -2.0 |
| 374.923 | -52.12 | 0.49 | 2583 | 188 | -19.96 | 1.15 | 43.6 | -6.4 |
| | | | | ($0.50\ CO_2 + 0.50\ O_2$) | | | | |
| 275.007 | -67.88 | 0.06 | 2747 | 8 | -51.13 | 0.60 | 23.9 | -0.8 |
| 299.946 | -54.62 | 0.02 | 2376 | 2 | -40.82 | 0.58 | 20.8 | -3.4 |
| 324.952 | -44.24 | 0.07 | 2132 | 11 | -32.75 | 0.60 | 19.6 | -4.9 |
| 349.935 | -35.79 | 0.21 | 1929 | 56 | -25.96 | 0.72 | 18.8 | -6.4 |
| 374.921 | -28.77 | 0.25 | 1778 | 74 | -20.21 | 0.77 | 18.6 | -8.2 |
| | | | | ($0.25\ CO_2 + 0.75\ O_2$) | | | | |
| 250.027 | -51.28 | 0.20 | 2030 | 41 | -63.39 | 0.71 | 10.2 | -4.7 |
| 260.026 | -46.82 | 0.05 | 1956 | 6 | -58.75 | 0.59 | 11.0 | -4.0 |
| 274.996 | -40.58 | 0.06 | 1827 | 8 | -51.65 | 0.59 | 10.5 | -4.4 |
| 299.942 | -31.90 | 0.09 | 1655 | 17 | -41.61 | 0.61 | 9.7 | -5.2 |
| 324.952 | -24.84 | 0.14 | 1523 | 32 | -33.36 | 0.65 | 9.1 | -6.0 |
| 349.938 | -19.16 | 0.14 | 1438 | 32 | -26.85 | 0.65 | 10.2 | -5.5 |
| 374.923 | -14.25 | 0.31 | 1343 | 98 | -20.98 | 0.86 | 10.1 | -6.4 |

**Table 8.** Parameters of the interpolation of the second interaction virial coefficient $B_{12}(T)$ for the binary ($CO_2$ + $O_2$) system as a function of temperature using Eq. (8)

| Parameter | Value ± expanded ($k$ = 2) uncertainty | Unit |
|---|---|---|
| $N_0$ | 68.8 ± 3.0 | $cm^3 \cdot mol^{-1}$ |
| $N_1$ | -33100 ± 820 | $cm^3 \cdot mol^{-1} \cdot K$ |
| RMS of residuals | 2.3 | % |

**Table 9.** Statistical analysis of the experimental ($p$, $\rho$, $T$) data set with respect to the GERG-2008 EoS (density residuals), EOS-CG (density residuals), and virial EoS (pressure residuals) for the five ($CO_2$ + $O_2$) mixtures studied in this work and the previous one [1], including literature data for comparable mixtures. AAD = absolute average deviation, Bias = average deviation, RMS = root mean square deviation, MaxD = maximum deviation.

| Reference[(a)] | $x(O_2)$ | $N$[(b)] | Covered ranges | | Experimental *vs* GERG-2008 | | | | Experimental *vs* EOS-CG | | | | Experimental *vs* virial EoS | |
|---|---|---|---|---|---|---|---|---|---|---|---|---|---|---|
| | | | $T$ / K | $p$ / MPa | AAD / % | Bias / % | RMS / % | MaxD / % | AAD / % | Bias / % | RMS / % | MaxD / % | RMS / % | MaxD / % |
| This study, part one [1] | 0.050321 | 45 | 275-375 | 0.5-8 | 0.29 | 0.28 | 0.45 | 1.5 | 0.077 | -0.0022 | 0.11 | 0.41 | 0.035 | 0.10 |
| This study, part one [1] | 0.099856 | 47 | 260-375 | 0.5-9 | 0.69 | 0.68 | 1.1 | 4.4 | 0.15 | 0.054 | 0.26 | 1.2 | 0.047 | 0.16 |
| This study, part one [1] | 0.199907 | 70 | 250-375 | 0.5-13 | 1.3 | 1.3 | 1.9 | 6.6 | 0.22 | -0.052 | 0.50 | 3.2 | 0.47 | 3.4 |
| This work | 0.499973 | 123 | 250-375 | 0.5-20 | 1.1 | 1.1 | 1.4 | 3.9 | 0.59 | -0.58 | 0.90 | 3.0 | 0.22 | 1.3 |
| This work | 0.749914 | 151 | 250-375 | 0.5-20 | 0.32 | 0.31 | 0.40 | 1.1 | 0.66 | -0.66 | 0.90 | 2.6 | 0.062 | 0.44 |
| Gururaja et al. [46] | 0.0-1.0 | 9 | 297-303 | 0.1 | 1.8 | -1.7 | 3.3 | 7.6 | 1.8 | -1.7 | 3.3 | 7.6 | 3.6 | 8.2 |
| Mantovanni et al. [40][(c)] | 0.060700 | 96 | 303-383 | 1-20 | 1.4 | -1.4 | 2.0 | 8.3 | 2.1 | -2.1 | 2.8 | 13 | 8.0 | 22 |
| Mantovanni et al. [40][c] | 0.129100 | 100 | 303-383 | 1-20 | 2.2 | -2.2 | 2.8 | 13 | 3.5 | -3.5 | 4.1 | 13 | 4.1 | 15 |
| Mazzoccoli et al. [45] | 0.044200 | 12 | 273-293 | 1-7 | 2.4 | 2.4 | 2.9 | 6.7 | 1.8 | 1.8 | 2.5 | 6.6 | 1.9 | 5.4 |
| Mazzoccoli et al. [45] | 0.148800 | 18 | 273-293 | 1-7 | 1.2 | 0.4 | 1.8 | 5.2 | 1.9 | -0.89 | 2.5 | 7.2 | 1.5 | 2.9 |
| Al-Siyabi [42] | 0.050000 | 26 | 323-423 | 8-40 | 1.3 | -1.1 | 1.5 | 3.0 | 1.5 | -1.5 | 1.7 | 3.2 | 9.1 | 19 |
| Commodore et al. [47] | 0.01246 | 112 | 324-400 | 2-35 | 0.17 | 0.094 | 0.26 | 1.2 | 0.15 | 0.0033 | 0.23 | 1.2 | 11 | 24 |
| Muirbrook [48] | 0.035-0.4 | 32 | 273.15 | Saturation | 43 | 38 | 75 | 247 | 39 | 34 | 70 | 235 | 22 | 44 |

[a] Only measurements in the vapour and supercritical phase have been considered.

[b] Number of experimental points.

[c] Used for the development of EOS-CG.